\documentclass{amia}
\usepackage{lipsum}
\usepackage{amsmath}
\usepackage{algpseudocode}
\usepackage{algorithm}
\usepackage{graphicx}
\usepackage{adjustbox}
\usepackage{color}
\usepackage{enumitem}
\usepackage{makecell}
\usepackage{tabularx}
\usepackage{booktabs} 
\usepackage{array}    
\usepackage{caption}  
\usepackage{hyperref} 
\usepackage{xcolor}
\hypersetup{
    colorlinks,
    linkcolor={red!50!black},
    citecolor={blue!10!black},
    urlcolor={blue!50!black}
}
\usepackage{paralist}
\usepackage{CJKutf8}
\usepackage{ulem}
\usepackage{longtable}
\usepackage{threeparttablex}

\setlist{topsep=0pt, leftmargin=*}

\begin{document}

\title{Advances in RNA secondary structure prediction and RNA modifications: Methods, data, and applications

}

\author{
Shu Yang$^{1,*}$, Nhat Truong Pham$^{2,*}$, Ziyang Li$^{3,*}$, Jae Young Baik$^{1,*}$, Joseph Lee$^{1}$, Tianhua Zhai$^{1}$, Weicheng Yu$^{1}$, Bojian Hou$^{1}$, Tianqi Shang$^{1}$, Weiqing He$^{1}$, Duy Duong-Tran$^{1,4,\dag}$, Mayur Naik$^{3,\dag}$, Li Shen$^{1,\dag}$}

\def\thefootnote{${*}$}\footnotetext{These authors contributed equally to this work.} 
\def\thefootnote{${\dag}$}\footnotetext{Corresponding authors: (DDT) duongtra@usna.edu; 121 Blake Rd, Annapolis, MD 21402, USA. (MN) mhnaik@cis.upenn.edu; 610 Levine Hall, 3330 Walnut St, Philadelphia, PA 19104, USA. (LS) li.shen@pennmedicine.upenn.edu; B306 Richards Building, 3700 Hamilton Walk, Philadelphia, PA 19104, USA.}

\institutes{
\small
    $^1$ Dept. of Biostatistics, Epidemiology and Informatics, University of Pennsylvania, Philadelphia, USA \\
    $^2$ Dept. of Integrative Biotechnology, Sungkyunkwan University, Suwon, Republic of Korea \\
    $^3$ Dept. of Computer and Information Science, University of Pennsylvania, Philadelphia, USA\\
    $^4$ Dept. of Mathematics, United States Naval Academy, Annapolis, USA  
}

\maketitle

\section*{Abstract}
\textit{Due to the hierarchical organization of RNA structures and their pivotal roles in fulfilling RNA functions, the formation of RNA secondary structure critically influences many biological processes and has thus been a crucial research topic.
This review sets out to explore the computational prediction of RNA secondary structure and its connections to RNA modifications, which have emerged as an active domain in recent years.
We first examine the progression of RNA secondary structure prediction methodology, focusing on a set of representative works categorized into thermodynamic, comparative, machine learning, and hybrid approaches.
Next, we survey the advances in RNA modifications and computational methods for identifying RNA modifications, focusing on the prominent modification types. Subsequently, we highlight the interplay between RNA modifications and secondary structures, emphasizing how modifications such as m6A dynamically affect RNA folding and vice versa. 
In addition, we also review relevant data sources and provide a discussion of current challenges and opportunities in the field.
Ultimately, we hope our review will be able to serve as a cornerstone to aid in the development of innovative methods for this emerging topic and foster therapeutic applications in the future.
}

\textbf{Keywords:} RNA secondary structures, RNA modifications, bioinformatics, machine learning, deep learning, RNA language models

\section{Introduction}
\label{sec:introduction}

Ribonucleic acids (RNAs) are crucial for fundamental biological processes, from catalyzing biochemical reactions to regulating gene expression in all organisms. Beyond their well-known role as intermediaries in the central dogma, RNAs exhibit extraordinary functional versatility. Ribozymes catalyze essential biochemical reactions, such as peptide bond formation in the ribosome, while riboswitches sense metabolites and regulate gene expression in response to environmental changes. Regulatory RNAs, including microRNAs and long non-coding RNAs, orchestrate complex cellular processes, ranging from development and differentiation to stress response and disease progression. Emerging classes, such as circular RNAs and RNA aptamers, continue to expand our understanding of RNA’s regulatory and structural repertoire. This functional diversity is underpinned by the structure of RNA, where its precise shape and structure enable it to carry out catalytic functions, bind to diverse molecules, and regulate complex cellular processes \cite{spitale2023probing,assmann2023rock,ganser2019roles, edwards2007riboswitches, fu2014non}. As a result, understanding RNA structure is not only fundamental to biological mechanisms but also holds promise for therapeutic innovation, including the development of RNA-based drugs and vaccines \cite{crooke2018rna}.

\begin{figure}[!tb]
  \begin{center}
    \includegraphics[width = \textwidth]{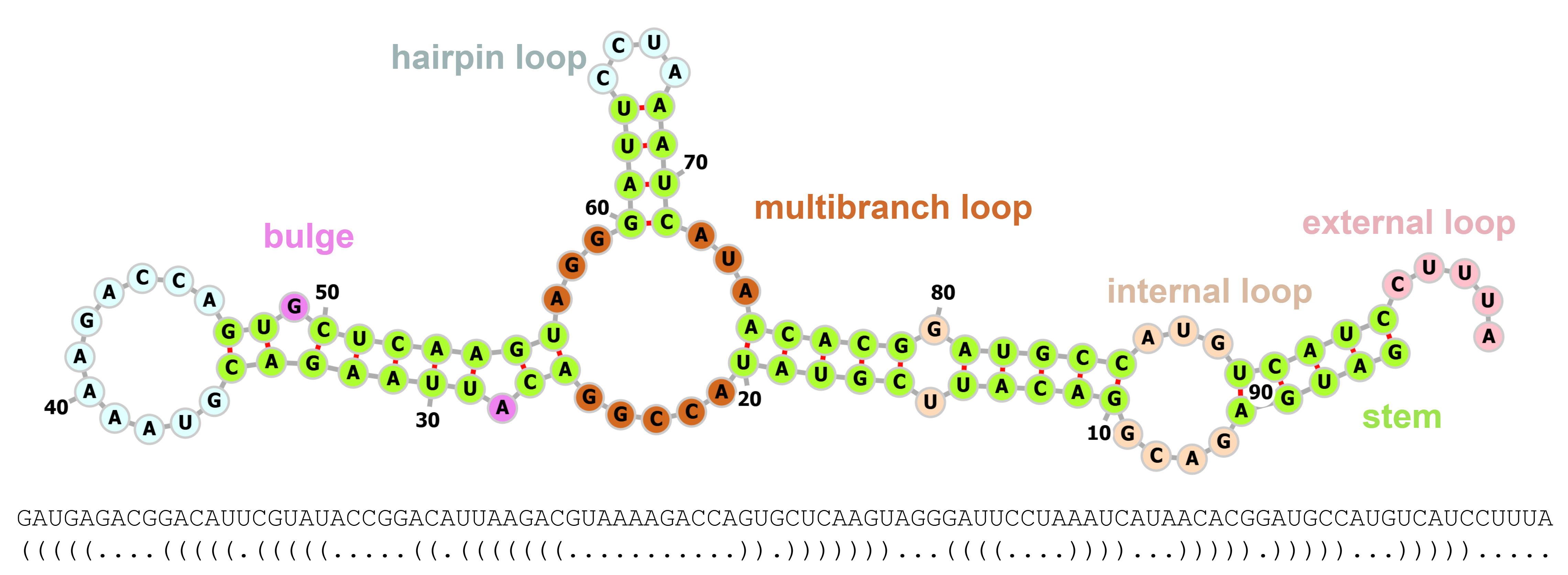}
    \caption{\textbf{An example of RNA secondary structures.} As shown in this example, an RSS can be decomposed into a set of stem-loop structural motifs (indicated with different colors). The RNA sequence and the corresponding dot-bracket notation are shown at the bottom. This plot is generated with the help of the Forna visualization tool from the ViennaRNA package~\cite{kerpedjiev2015forna}. The example assumes pseudoknot-free.}\label{fig:RSS} 
  \end{center}
\end{figure}

The structure of RNA exhibits a hierarchical and sequential organization: the primary structure consists of a linear sequence of nucleotides, the secondary structure such as stems and loops is dictated by base pair interactions, the tertiary structure builds upon these shapes with three-dimensional folding, and this determines quaternary interactions with other molecules \cite{nowakowski1997rna,tinoco1999rna, jones2015rna}. 
Among these, the secondary structure holds particular significance as it serves as a critical scaffold for higher-order folding and governs key aspects of RNA function. 
RNA secondary structure (RSS) formation is driven by the ability of nucleotides to engage in both canonical Watson–Crick basepairings and noncanonical ones, such as GU pairs. 
These pairings give rise to distinct secondary structural motifs (see Figure~\ref{fig:RSS} for an example), including stems, bulges, different types of loops, and pseudoknots, which consist of various configurations of paired and unpaired nucleotides. 
These secondary elements further interact with each other through mechanisms like coaxial stacking and kissing loop interactions, ultimately assembling into the more complex tertiary or quaternary structures of RNA.
Notably, RNA secondary structure is often more conserved than its primary sequence across homologous RNAs, emphasizing its central role in defining RNA behavior and function \cite{mathews2010folding}. 

Determining RNA structures through experimental techniques (e.g., X-ray crystallography) remains a significant challenge due to their low throughput, high resource demands, and technical limitations. These methods have resolved only a fraction of known RNA molecules, leaving significant gaps in our understanding of RNA structure. 
To address these limitations, computational prediction has emerged as an essential tool for scaling RNA structural analysis. However, while protein folding has seen transformative advancements, such as AlphaFold’s ability to predict 3D structures with remarkable accuracy \cite{jumper2021highly}, RNA folding remains a difficult challenge. Firstly, 
one fundamental obstacle in RNA structure prediction is the inadequacy and bias of existing datasets. 
While protein structure prediction has benefited from large, high-quality databases, RNA data in repositories like the Protein Data Bank (PDB) is both limited and heavily skewed towards simpler structures, such as tRNAs and rRNA subunits. 
For example, there are 25 times more proteins structure data entries than RNAs in PDB.
There are $>$19,600 protein families in Pfam, but only $>$4,100 RNA families in Rfam.
Moreover, the lack of structural diversity also hinders the development of models that can accurately predict the wide variety of RNA structural motifs, e.g. long-range interactions in lncRNAs. 
In addition, unlike proteins, whose folding is driven by well-characterized forces like hydrophobic interactions and standardized motifs, RNA folding is more dynamic and involves intricate secondary structures that form the scaffold for complex tertiary interactions. 
This makes secondary structure prediction an essential focus in RNA research.

Over the years, three major computational prediction strategies have emerged for RSS folding: thermodynamic, comparative, and machine learning (including deep learning). 
Thermodynamic models predict structures by minimizing free energy, using experimentally derived parameters to estimate the stability of base-pair interactions and loops. 
While effective for canonical structures, these methods are limited by incomplete energy models and challenges in handling non-canonical pairs and pseudoknots. 
Comparative approaches leverage evolutionary conservation to predict RNA secondary structures, assuming that functionally important structures are preserved through compensatory mutations. 
These methods typically rely on a multiple sequence alignment (MSA) of homologous RNA sequences and use probabilistic models, such as stochastic context-free grammars (SCFGs), to identify co-varying base pairs. 
Instead of minimizing energy, they predict the structure that best explains the observed evolutionary covariation, selecting the one with the highest likelihood. 
Machine learning and deep learning methods, in contrast, learn structural patterns in a data-driven way, capturing complex interactions without explicit reliance on energy models or sequence homology. These models, exemplified by SPOT-RNA \cite{singh2019rna} and Ufold\cite{fu2022ufold}, offer state-of-the-art accuracy for diverse RNA structures, including those with intricate motifs and pseudoknots, marking a paradigm shift in RNA structure prediction. 
Notably, foundational models like RNA-FM\cite{chen2022interpretable} have further advanced the field, leveraging vast amounts of RNA sequences with unsupervised learning.

Accurate prediction of RSS from the RNA sequence is fundamental to understanding RNA function and interactions; however, there are many other factors affecting RNA folding in the cell. 
Among them, chemical modifications of RNA, such as methylation and pseudouridylation, play pivotal roles in influencing RNA thermodynamic stability by altering the free energy landscape. These modifications can impact base pairing, reshape the secondary structure, and modulate RNA interactions with proteins and other biomolecules, thereby influencing key biological processes such as translation, splicing, and RNA stability.
Indeed, the so-called epitranscriptome of RNA modifications is suggested to have broad regulation of RNA structuredness, supported by observations of transcriptome wide RNA modifications~\cite{gilbert2016messenger, tanzer2019rna}.
Notably, RNA modifications are remarkably diverse and widely distributed across various RNA species. While covalent nucleotide modifications were traditionally recognized as abundant in transfer RNAs (tRNAs), advances in high-throughput sequencing technologies have revealed their widespread occurrence in messenger RNAs (mRNAs) and non-coding RNAs. As of this review, the Modomics database catalogs over 335 natural RNA modifications~\cite{cappannini2024modomics}, though most remain incompletely characterized. Emerging evidence underscores the critical roles of these modifications in shaping RNA secondary and tertiary structures~\cite{tanzer2019rna, kierzek2022secondary}, facilitating RNA-protein interactions~\cite{lewis2017rna, boo2020emerging}, and regulating splicing~\cite{wang2022dynamic}. A comprehensive understanding of the interplay between RNA modifications and structural dynamics is crucial for elucidating RNA biology, with far-reaching implications for regulatory mechanisms and therapeutic development.

Specifically, post-transcriptional RNA modifications, which involve chemical alterations to RNA molecules, have recently emerged as a major focus of research~\cite{cui2022rna, liu2024rna, ramos2022rna}. Over the past decade, several prominent modifications have been extensively explored, including N6-methyladenosine (m6A), N1-methyladenosine (m1A), 5-methylcytosine (m5C), 5-methyluridine (m5U), N6,2'-O-dimethyladenosine (m6Am), N7-methylguanosine (m7G), N4-acetylcytosine (ac4C), pseudouridine ($\Psi$), 2'-O-methyladenosine (Am), 2'-O-methylcytidine (Cm), 2'-O-methylguanosine (Gm), 2'-O-methyluridine (Um), uridylation, and adenosine-to-inosine (A-to-I) RNA editing~\cite{song2021attention, qiao2024towards, chen2023transrnam}. Among these, m6A is recognized as the most prevalent and abundant mRNA modification in eukaryotes, accounting for approximately 0.1–0.6\% of all adenosines. This modification involves the addition of a methyl group to the nitrogen atom at the sixth position of adenosine and is conserved across a wide range of organisms, from bacteria to mammals. Another well-studied modification, m5C, is characterized by the addition of a methyl group to the carbon-5 position of the cytosine base. This modification is commonly observed across various RNA species, including tRNAs, ribosomal RNAs (rRNAs), mRNAs, enhancer RNAs (eRNAs), and non-coding RNAs. Additionally, 2'-O-methylation (Nm or 2OM) is an RNA modification that occurs co-transcriptionally or post-transcriptionally, involving the addition of a methyl group to the 2' hydroxyl group of the ribose sugar in the RNA backbone. These modifications collectively contribute to the structural and functional diversity of RNA molecules, enabling their involvement in a broad spectrum of biological processes.

Subsequently, several experimental methods have been developed to accurately identify modifications in RNA. However, these methods tend to be labor-intensive, require specialized tools, can damage RNA samples, and present challenges when working with minimal RNA quantities. To address these limitations, high-throughput techniques based on deep sequencing have been introduced to identify RNA modifications at the transcriptome level. Nevertheless, these techniques remain costly, time-consuming, and necessitate specialized expertise. Consequently, computational methods have been developed to complement the experimental techniques. Recently, several databases have been established, serving as foundational resources for developing computational methods to identify RNA modifications, such as RMBase V3.0~\cite{xuan2024rmbase}, MODOMICS~\cite{cappannini2024modomics}, m7GHub V2.0~\cite{wang2024m7ghub}, and the dataset by Song~\textit{et~al.}~\cite{song2021attention}. Furthermore, a variety of computational tools have emerged for several common RNA modifications, including 2OM (H2Opred~\cite{pham2024h2opred}, Meta-2OM~\cite{harun2024meta}, and Nmix~\cite{geng2024nmix}), ac4C (ac4C-AFL~\cite{pham2024ac4c}, Voting-ac4C~\cite{jia2024voting}, iRNA-ac4C~\cite{su2023irna}, and TransAC4C~\cite{liu2024transac4c}), m5C (Deepm5C~\cite{hasan2022deepm5c}, MLm5C~\cite{kurata2024mlm5c}, and m5C-pred~\cite{abbas2023xgboost}), m6A (MST-m6A~\cite{su2024mst}, CLSM6A, and BLAM6A-Merge~\cite{xia2024blam6a}), m7G (Moss-m7G~\cite{zhao2024moss} and THRONE~\cite{shoombuatong2022throne}), and multiple transcriptome modifications (MultiRM~\cite{song2021attention}, TransRNAm~\cite{chen2023transrnam}, and class incremental learning for RNA modifications (CIL-RNA)~\cite{qiao2024towards}). Detailed information on these tools will be discussed in later sections.

Here, this review aims to survey the advances in RNA secondary structure prediction and RNA modification, with an emphasis on the connections between these two, and to outline future directions for this emerging field.
There are a number of existing reviews on related topics recently.~\cite{bugnon2022secondary, sato2023recent, schneider2023will, wu2023machine, zhang2022advances} 
Our review sets itself apart by: 1. focusing on a representative set of diverse RSS prediction methods to offer more in-depth descriptions for each, rather than giving a broad but brief overview, and to show the methodology progression in this field, rather than concentrating on specific types of methodology (e.g., deep learning), so that the review could be hopefully beneficial for researchers or practitioners relatively familiar with the topic and also useful for those with non-expert backgrounds; and 2. dedicating substantial efforts to discussing the potential applications of RNA secondary structure prediction in the context of RNA modifications, a cutting-edge topic with immense potentials in human disease study. 
In the following sections of the paper, we will first review the progression of RNA secondary structure prediction methods by concentrating on a set of selected works representing different strategies in the field. 
Then, we will shift our focus to RNA modification prediction methods with another set of representative works corresponding to different modification types. 
After that, we will delve into the interplay between RNA secondary structure and RNA modification, building on top of the earlier sections to highlight the close relationship between the two.
Additionally, we will provide a summary of available data that are commonly used in this area.
Lastly, we will discuss the challenges and opportunities for future work to conclude the review.

\section{Computational tools for RNA secondary structure prediction} 
\label{sec:RSS}

As mentioned above, RNA structure is formed hierarchically, and the secondary structure formation is key to study the functions of the RNA. 
Thus, the in silico RNA secondary structure prediction has long been a cornerstone in bioinformatics. 
The first fast algorithm for RSS prediction was published in 1980 \cite{Nussinov1980} by Nussinov and Jacobson, a foundational contribution that continues to influence the field even now.
After that, the field has evolved significantly over the past decades, especially in recent years due to the success in deep learning and large foundation models.
Therefore, we select a number of representative methods to review here, reflecting the progression of methodology developments and hopefully shedding lights on potential future directions. 

RNA secondary structure prediction methods can be classified into different categories based on different but subtly related criteria.
Conventional, the most commonly seen classifications being the energy-based model versus the probability model (e.g., stochastic context-free grammar model), according to the type of parameters used, or single-sequence structure prediction versus comparative structure prediction, according to the type of inputs required.
In a loose sense, energy-based method roughly overlaps with single-sequence structure prediction, while the probability model largely overlaps with comparative prediction.
Recent advancements of machine learning- and deep learning-based approaches have drastically changed the research paradigm.
In addition, the emerging hybrid methods have made the boundaries among different classes become even more vague.
So here, in order to show the methodology progression in this field, we follow a straightforward notion to refer to these methods as energy-based, comparative, learning-based, and hybrid methods.
A summary of the key ideas of the different classes are shown in Table~\ref{MFEvsCmp}. 
Besides, technically speaking, SCFG is also learning-based as a probabilistic model itself; but since it has been widely-used in comparative methods, we single it out without putting it together with the other machine learning or deep learning methods.

\begin{table}[t]
\caption{
\textbf{A summary of RNA secondary structure prediction strategies.}}
\begin{center}
\begin{threeparttable}
\begin{tabular}{lp{0.48\textwidth}p{0.25\textwidth}}
\toprule
\textbf{Strategy} & \textbf{Method} & \textbf{Input} 
\\\midrule
\textbf{Energy-based} & Based on thermodynamic free energy, find an optimal structure with minimum free energy & Single sequence  \\
\\
\textbf{Comparative} & Based on co-variation and probabilistic model, find the optimal structure with maximum likelihood & Multiple sequence alignment  \\
\\
\textbf{Learning-based} \tnote{1} & Data driven, advanced machine learning or deep learning approaches & Diverse   \\
\\
\textbf{Hybrid} & Combining two or more strategies mentioned above  & Diverse  \\
\bottomrule
\end{tabular}
\begin{tablenotes}
    \item[1] Learning-based strategy here refers to those non-SCFG-based machine learning and deep learning methods since, technically speaking, the classic SCFG model is also learning-based.
\end{tablenotes}
\end{threeparttable}
\\[10pt]
\label{MFEvsCmp} 
\end{center}
\end{table}

\subsection{Thermodynamic free energy-based methods}
    
\begin{figure}[!tb]
  \begin{center}
    \includegraphics[width = 0.9\textwidth]{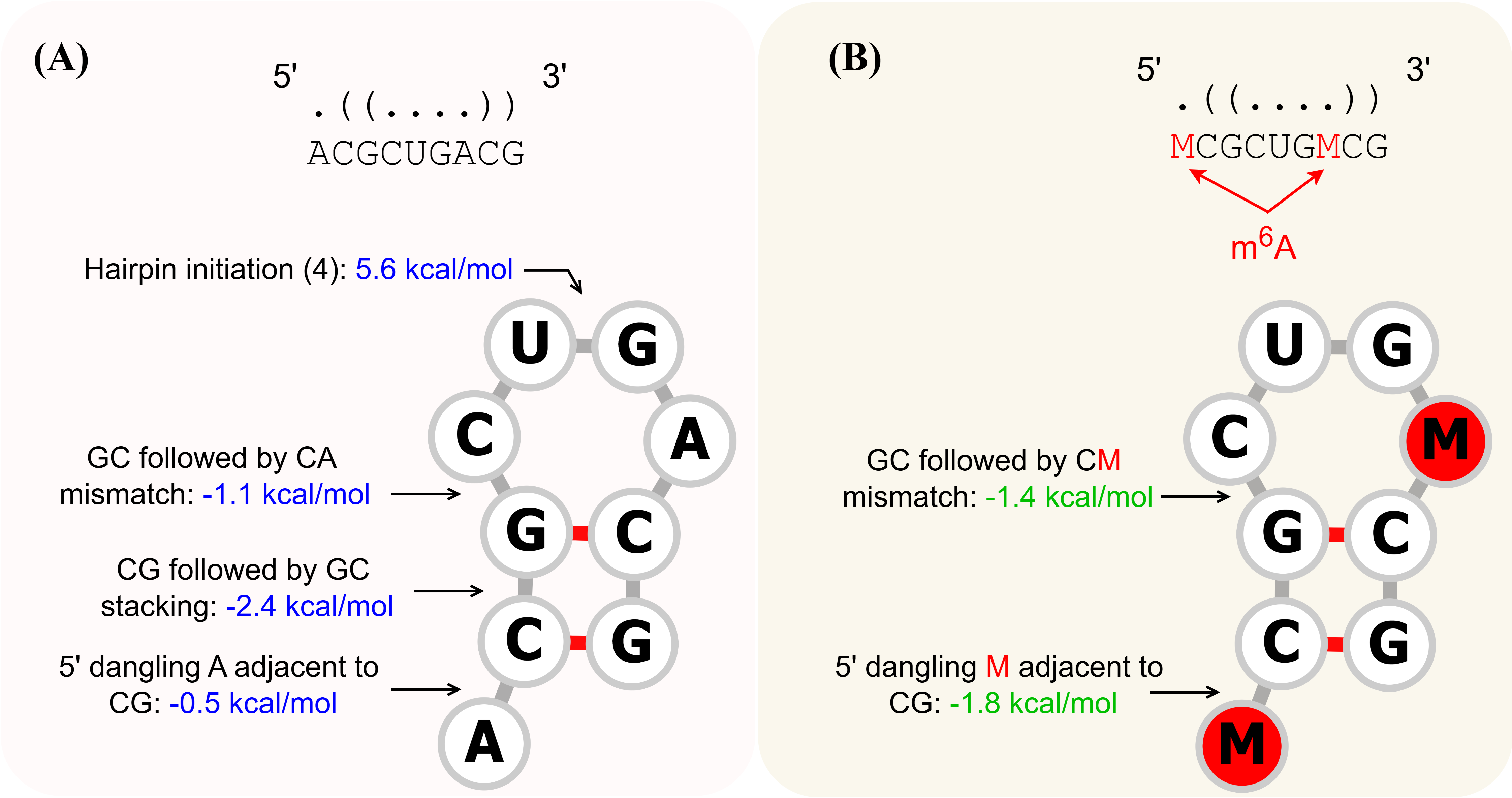}
    \caption{\textbf{RSS free energy computation based on nearest neighbor energy model.} \textbf{(A).} An example of the free energy calculation for an RSS using Turner's nearest neighbor parameters~\cite{mathews2004incorporating}. 
    This figure is generated with the help of the Forna visualization tool from the ViennaRNA package~\cite{kerpedjiev2015forna}. 
    The overall free energy of a given RNA secondary structure can be expressed as the sum of free energies across different structural units. 
    For this small hairpin structure shown here, the overall free energy is the sum of the destabilizing loop and bulge energy (e.g. [+5.6] for hairpin initiation (4) as shown in the figure) and the stabilizing energy contributions from pairs of neighboring basepairs (e.g. [-2.4] for the CG followed by GC stacking interaction).
    \textbf{(B).} A side-by-side comparison of m6A modified RNA with a different nearest neighbor model~\cite{kierzek2022secondary}, assuming the same RSS as the unmodified one.
    As we can see, the energy values are very different for structures involving the normal A from those involving m6A (denoted by letter M in the figure).
    For example, the 5' dangling A has [-0.5] while m6A has [-1.8].}\label{MFEfig} 
  \end{center}
\end{figure}

The idea of energy-based structure prediction is based on the principle of free energy minimization. Since the secondary structure of RNA molecules is predominantly determined by interactions like hydrogen bonds and base stackings, computing the energy of these interactions provides insights into the structure.
For a closed system with fixed entropy, equilibrium corresponds to a state that minimizes the system’s free energy. Therefore, the most stable secondary structure of RNA is assumed to be the one with the minimum free energy (MFE).
Methods in this category predict the secondary structure that minimizes the overall free energy of an RNA in thermodynamic equilibrium by considering all potential RNA secondary structures and their respective abundance according to the Boltzmann distribution.

The energy-based structure prediction typically takes a single RNA sequence as input and predicts the best structure of that sequence to be the most stable structure, i.e. the one with the minimum free energy.
Such prediction uses the stacking of base pairs as its basic unit. 
As shown in Figure~\ref{fig:RSS}, an RNA secondary structure can be decomposed into a set of stem-loop structural motifs (indicated with different colors) such as stacking/stem, hairpin loop, internal loop, bulge, multibranch loop, and external loop/dangling region, etc.
The prediction method should have parameters to account for all these motifs, in terms of their energetic contributions.

The nearest neighbor models~\cite{schroeder2009optical, turner2010nndb} provide the thermodynamic free energy parameters and are the basis for most methods involving energy-based computations. 
The nearest neighbor models are a set of rules and associated parameters that predict the folding free energy of a secondary structure by decomposing the structure into loop substructures enclosed by the nearest neighboring basepairs. 
The model rules have two general assumptions: 1. the free energy of a basepair or loop substructure depends only on the sequence of that substructure and the sequence of the directly adjacent basepairs. 2. the total free energy can be calculated by summing the energies predicted for each substructure. 
The set of model parameters stores the energy values for the smallest unit of secondary structure, corresponding to thermodynamic quantities pre-determined by wet lab experiments such as optical melting experiments. 
The most famous one is the Turner's nearest neighbor model, which has been widely used by many methods including those energy-based methods in Table~\ref{CMPs}. 

Figure~\ref{MFEfig}\textbf{(A)} shows an example of the nearest neighbor model of an RNA structure. 
The additivity characteristic indicates that globally optimal structures are composed of locally optimal sub-structures, which is ideal for computational approaches like dynamic programming algorithms to deal with. 
Dynamic programming (DP) algorithm is the first and most widely used approach for energy-based RSS prediction \cite{Nussinov1980,zuker1981optimal,Stadler2012}. 
It can recursively calculate the minimum energy structure.
As the goal in RNA structure prediction is to find the RNA structure that minimizes the overall free energy for a given RNA input sequence, and as the overall free energy can be expressed as the sum of free energy contributions from structural building blocks recursively, efficient DP algorithms exist to calculate the optimal RNA secondary structure in $O(N^3)$ time and $O(N^2)$ memory for an input sequence of length $N$.

Besides the structural motifs shown in Figure~\ref{fig:RSS}, additional structures can be formed when unpaired bases match with distant ones, such as the base A at position 40 and the U at position 6 in Figure~\ref{fig:RSS}. 
This would form the so-called pseudoknot structure, which could make the prediction much harder.
For example, the Zuker algorithm~\cite{zuker1981optimal} underlying many DP algorithms like RNAfold~\cite{lorenz2011viennarna} etc. is incapable of predicting pseudoknot; while a later enhanced DP algorithm by Rivas and Eddy to deal with pseudoknots has time complexity of $O(N^6)$ and space complexity $O(N^4)$~\cite{rivas1999dynamic}, which is often intractable in practice.
Note that since many of the methods reviewed below do not deal with pseudoknots, we will specify this capability if otherwise for clarity.

\paragraph{\href{http://rna.tbi.univie.ac.at/cgi-bin/RNAWebSuite/RNAfold.cgi}{\textsc{RNAfold}}} One of the pioneers and most widely used MFE-based RNA secondary structure prediction methods is \textsc{RNAfold} from the ViennaRNA package~\cite{lorenz2011viennarna}.
It employs Zuker’s dynamic programming algorithm~\cite{zuker1981optimal} for efficient computation of the optimal global folding with Turner's nearest neighbor free energy parameters as the scores~\cite{mathews288expanded, mathews2004incorporating}. 
By accepting a single RNA sequence in FASTA format as input, RNAfold systematically evaluates possible base-pairing interactions to identify the thermodynamically most stable secondary structure, simultaneously reporting the predicted MFE values and providing dot-bracket notations for quick visualization. 
Zuker's algorithm has a time complexity of $O(N^3)$ that scales cubically with the sequence length $N$.
RNAfold also has the functionality to output the MFE probability and base-pairing probability matrix by utilizing McCaskill’s partition function approach~\cite{mccaskill1990equilibrium} (also $O(N^3)$ runtime but with a larger constant factor than the Zuker's) to consider the thermodynamic ensemble of all structures following the Boltzmann distribution.
RNAfold is based on well-established models in the early days and serves as the main program in the ViennRNA package, making it one of the most classic and reliable MFE-based methods for RNA secondary structure prediction. 

\paragraph{\href{https://rna.urmc.rochester.edu/RNAstructure.html}{\textsc{RNAstructure}}} \textsc{RNAstructure} is a software tool for RNA secondary structure prediction and analysis~\cite{reuter2010rnastructure, Ali2023RNAstructure}. The algorithms in \textsc{RNAstructure} employ nearest neighbor parameters to predict the stability of secondary structures. These parameters, based on the Turner group~\cite{xia1998thermodynamic, mathews2004incorporating, lu2006set}, include both free energy change at 37$^{\circ}$C and enthalpy change to facilitate the prediction of conformational stability at various temperatures. \textsc{RNAstructure} offers a range of algorithms, including methods for secondary structure prediction, base pair probability estimation, bimolecular structure prediction, and identifying common structures between two sequences. These features are complemented by a user-friendly JAVA-based GUI with cross-platform compatibility.

\paragraph{\href{https://www.cs.ubc.ca/labs/algorithms/Projects/RNA-Params/}{\textsc{SimFold}}} \textsc{SimFold} is a computational tool designed to predict RNA secondary structures using thermodynamic free-energy models. It employs advanced parameter estimation techniques, such as the Constraint Generation (CG) and Boltzmann Likelihood (BL) methods, to optimize energy parameters for RNA folding. \textsc{SimFold} supports the Turner energy model and its variants, effectively integrating structural and thermodynamic data to enhance prediction accuracy. Benchmark studies demonstrate that \textsc{SimFold} achieves a significant improvement in F-measure over standard Turner parameters, particularly when optimized with the BL method. The tool accepts input sequences in FASTA format and outputs RNA secondary structures with minimum free energy, offering accurate and reliable predictions for pseudoknot-free RNA configurations. \textsc{SimFold}’s robust thermodynamic modeling and energy parameter refinements make it an essential tool for RNA structure prediction tasks.

\paragraph{\href{https://linearfold.eecs.oregonstate.edu/}{\textsc{LinearFold}}} \textsc{LinearFold} is the first linear-time and linear-space prediction algorithm for RNA secondary structures. It uses the thermodynamic free energy model \cite{mathews2004incorporating} from Vienna RNAfold \cite{lorenz2011viennarna}. Traditional algorithms for predicting RNA structures, such as dynamic programming-based methods, scale with cubic time complexity $O(N^3)$, which limits their use for long RNA sequences. \textsc{LinearFold} overcomes this bottleneck by scanning the sequence in a left-to-right (5'-to-3') direction, following the transcription process, rather than bottom-up, utilizing a beam search heuristic to reduce the complexity to $O(N)$, making it significantly faster without a large sacrifice in accuracy. It accepts input sequences in both FASTA format and pure-sequence format. \textsc{LinearFold} demonstrates superior efficiency and scalability without sacrificing accuracy, making it particularly effective for long sequences and long-range base pair predictions.

\subsection{Comparative methods}

The comparative strategy usually uses probabilities as parameters and stochastic context free grammars as the underlying model.
Although there are exceptions~\cite{rivas2012tornado}, it typically needs a functionally equivalent multiple sequence alignment as input and predicts the best structure of that alignment to be the most likely structure, e.g., the one with the maximum likelihood. 

\begin{figure}[!tb]
  \begin{center}
    \includegraphics[width = 0.7\textwidth]{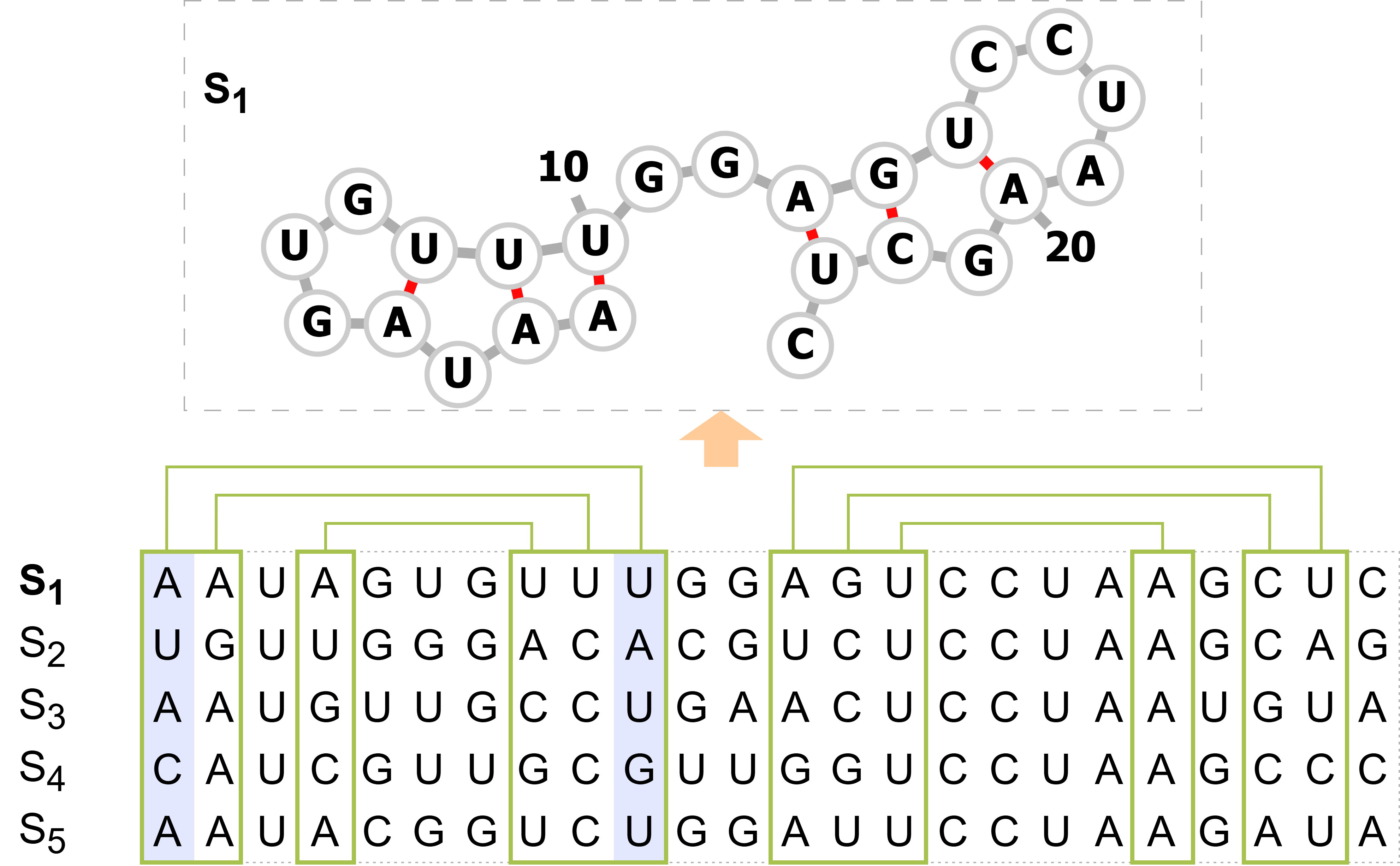}
    \caption{\textbf{A schematic illustration of comparative RNA secondary structure prediction strategy.} Assuming the functionally important structures (not necessarily the sequences) are conserved through evolution, the idea is to first align equivalent sequences from related species ($\mathrm{S}_1, ..,\mathrm{S}_5$), then find pairs of co-varying alignment columns (green), finally incorporate the co-varying information into the predicted structure.}\label{CMPfig} 
  \end{center}
\end{figure}

The idea of the comparative structure prediction is based on evolution, as shown in Figure \ref{CMPfig}.
It assumes functionally important RNA structures are conserved through evolution.
So, it looks for conserved base pairs, especially those compensatory mutations through evolution. 
An algorithm in this category needs the alignment of functionally equivalent sequences, and it finds those co-varying base pairs in the alignment columns. 
Instead of using energy parameters, it uses a probabilistic model like SCFG. 
The final predicted RNA structure would be the one that best explains those co-variations according to the model.

\begin{ThreePartTable}
    \begin{TableNotes}\footnotesize
        \item \textbf{Note:} Methods are sorted by their categories (\textbf{E}: energy-based methods. \textbf{C}: comparative methods. \textbf{L}: learning-based methods. \textbf{H}: hybrid methods) and alphabet order. MFE: Minimum free energy. SCFG: Stochastic Context-Free Grammar. PSSM: Position Specific Score Matrix. DCA: Direct Coupling Analysis
    \end{TableNotes}
    
    \begin{small}    
    \begin{longtable}{lp{0.23\textwidth}p{0.17\textwidth}p{0.23\textwidth}p{0.09\textwidth}}
    \caption{ \textbf{A summary of representative RNA secondary structure prediction methods included in the review.} }
    \label{CMPs}\\
    \hline
    \textbf{} & \textbf{Method} & \textbf{Input} & \textbf{Other notes} & \textbf{URLs} \\
    \hline
    \endfirsthead
    \multicolumn{5}{l}{Continuation of table}\\
    \hline
    \textbf{} & \textbf{Method} & \textbf{Input} & \textbf{Other notes} & \textbf{URLs} \\
    \hline
    \endhead
    \hline
    \multicolumn{5}{r}{More on next page}\\
    \endfoot
    \hline
    \insertTableNotes
    \endlastfoot
    \textbf{{\sc LinearFold}}~\cite{huang2019linearfold} & \textbf{(E)}. Incremental Beam Search algorithm with a thermodynamic energy model & Single sequence (FASTA) & Sequence length limits to 100,000  & \href{https://github.com/LinearFold/LinearFold}{Code}, \href{https://linearfold.eecs.oregonstate.edu/}{Web server}\\
    \\
    \textbf{{\sc RNAfold}}~\cite{lorenz2011viennarna} & \textbf{(E)}. MFE with Zuker's and McCaskill's algorithms & Single sequence (FASTA) & One of the most classic MFE methods and the main RSS prediction tool in ViennaRNA & \href{https://www.tbi.univie.ac.at/RNA/RNAfold.1.html}{Code}, \href{http://rna.tbi.univie.ac.at/cgi-bin/RNAWebSuite/RNAfold.cgi}{Web server}\\
    \\
    \textbf{{\sc RNAstructure}}~\cite{reuter2010rnastructure, Ali2023RNAstructure} & \textbf{(E)}. Software with a set of folding algorithms  & Single or multi-sequence (FASTA, SEQ) & Predicts MFE structures, base pair probabilities, bimolecular structures, and structures common to two sequences  & \href{https://rna.urmc.rochester.edu/RNAstructureDownload.html}{Code}, \href{https://rna.urmc.rochester.edu/RNAstructureWeb/}{Web server}\\
    \\
    \textbf{{\sc SimFold}}~\cite{andronescu2003algorithms, andronescu2007efficient} & \textbf{(E)}. Minimum free energy based on a discriminative framework & Single sequence (FASTA) & Current implementation includes suboptimal folding calculations, as well as partition functions, base pair probabilities, and gradient computations.  & \href{https://www.cs.ubc.ca/labs/algorithms/Projects/RNA-Params/}{Code}\\
    \\
    \textbf{{\sc PPfold}}~\cite{sukosd2012ppfold, sukosd2011multithreaded} & \textbf{(C)}. SCFG with phylogenetics information & Alignment; phylogenetic tree (optional) & Uses probabilities as parameters to measure the co-varying tendency of positions pair & \href{http://birc.au.dk/software/ppfold}{Code}\\
    \\
    \textbf{{\sc RNAdecoder}}~\cite{pedersen2004comparative, pedersen2004evolutionary} & \textbf{(C)}. SCFG with several phylogenetic models & Alignment; phylogenetic tree; codon annotation & Designed explicitly to take protein-coding context into account; does not assume global RNA structure & \href{https://e-rna.org/rnadecoder/download/rnadecoder.tar.bz2}{Code}, \href{https://e-rna.org/rnadecoder/}{Web server}\\
    \\
    \textbf{{\sc TORNADO}}~\cite{rivas2012tornado} & \textbf{(C)}. Generalized super-grammar for SCFG described in TORNADO programming language & Single sequence (FASTA, Stockholm) & General-purpose SCFG tool with flexible adaptability; supports complex structural modeling and customization & \href{https://github.com/EddyRivasLab/tornado}{Code}\\
    \\
    \textbf{{\sc CNNFold}}~\cite{saman2022rna} & \textbf{(L)}. Deep learning-based model using CNN architecture & Single sequence (FASTA) & Simple implementation of CNN-based method & \href{https://github.com/mehdi1902/RNA-secondary-structure-prediction-using-CNN}{Code}\\
    \\
    \textbf{{\sc CONTRAfold}}~\cite{Do2006contrafold} & \textbf{(L)}. Conditional log-linear models (CLLMs), a discriminative generalization of SCFG & Single sequence (FASTA) & Construct the CLLM from the energy model to find the maximum-expected-accuracy structure. Sequence length limits to 1000 on the web server. & \href{http://contra.stanford.edu/contrafold/download.html}{Code}, \href{http://contra.stanford.edu/contrafold/server.html}{Web server}
    \\
    \textbf{{\sc E2Efold}}~\cite{Chen2020RNA} & \textbf{(L)}. End2end learning with a transformer-based Deep Score Network and a multilayer Post-Processing Network with an unrolled algorithm to reduce overfitting & One-hot encoding of single sequence & The unrolled algorithm uses a primal-dual constrained optimization to incorporate base pairing constraints &  \href{https://github.com/ml4bio/e2efold}{Code}\\
    \\        
    \textbf{{\sc REDfold}}~\cite{chen2023redfold} & \textbf{(L)}. Deep learning-based model using ResNet and FC-DenseNet network& Single sequence (FASTA) &  &  \href{https://github.com/aky3100/REDfold}{Code}, \href{https://redfold.ee.ncyu.edu.tw/redfold/}{Web server}\\
        \\
    \textbf{{\sc SPOT-RNA}}~\cite{singh2019rna} & \textbf{(L)}. Ensemble of ResNets, 2D-BiLSTM and dilated CNN models; transfer learning & One-hot encoded single sequence (or batch sequences) & Able to predict all base pairs including noncanonical and non-nested (pseudoknot) ones & \href{https://github.com/jaswindersingh2/SPOT-RNA}{Code}, \href{https://sparks-lab.org/server/spot-rna/}{Web server}
    \\ 
    \textbf{{\sc SPOT-RNA2}}~\cite{singh2021improved} & \textbf{(L)}. Ensemble of dilated CNN models; transfer learning & One-hot encoding and predicted basepair probability from single sequence, PSSM and DCA from evolution & Extends SPOT-RNA by  incorporating additional evolutionary information &  \href{https://github.com/jaswindersingh2/SPOT-RNA2}{Code}, \href{https://sparks-lab.org/server/spot-rna2/}{Web server}
    \\
    \textbf{{\sc UFold}}~\cite{fu2022ufold} & \textbf{(L)}. Deep learning-based model using U-Net architecture & Single sequence (FASTA) & One of the first deep-learning models that converts RNA sequences to an ``image" format & \href{https://github.com/uci-cbcl/UFold}{Code}\\
    \\
    \textbf{{\sc MxFold}}~\cite{akiyama2017mxfold} & \textbf{(H)}. Thermodynamic and structured support vector machines hybrid & Single sequence (FASTA, bpseq) & Integrates thermodynamic parameters with machine learning to improve prediction accuracy; limited scalability for very long sequences & \href{https://github.com/mxfold/mxfold}{Code}, \href{http://www.dna.bio.keio.ac.jp/mxfold/}{Web server}\\
    \\  
    \textbf{{\sc MxFold2}}~\cite{sato2021rna} & \textbf{(H)}. Thermodynamic and deep learning hybrid & Single sequence (FASTA, bpseq) & Integrates thermodynamic parameters with CNN, BiLSTM to leverage the power of deep models & \href{https://github.com/mxfold/mxfold2}{Code}, \href{https://ws.sato-lab.org/mxfold2/}{Web server}\\
    \\  
    \textbf{{\sc RNAalifold}}~\cite{bernhart2008rnaalifold} & \textbf{(H)}. MFE and covariation & Alignment & In contrast to {\sc Pfold}, uses free energies as parameters; modifies the scoring scheme of conventional MFE based dynamic programming algorithm  & \href{https://www.tbi.univie.ac.at/RNA/RNAalifold.1.html}{Code}, \href{http://rna.tbi.univie.ac.at/cgi-bin/RNAWebSuite/RNAalifold.cgi}{Web server}\\
    \\ 
    \textsc{CentroidFold}~\cite{sato2009centroidfold} & \textbf{(H)}. $\gamma$-centroid estimator & Single sequence (FASTA); multiple alignment (CLUSTAL)& Sequence length limits to 400  & \href{https://github.com/satoken/centroid-rna-package}{Code}, \href{http://rtools.cbrc.jp/centroidfold/}{Web server}\\
    \\
    \textbf{{\sc RNAErnie}}~\cite{wang2024multi} & \textbf{(H)}. Pre-trained, foundational model using the transformer architecture & Single sequence (FASTA) & Masks tokens on various semantic levels (e.g. RNA motifs) to encode richer biological information & \href{https://github.com/CatIIIIIIII/RNAErnie}{Code}\\
    \\    
    \textbf{{\sc RNA-FM}}~\cite{chen2022interpretable} & \textbf{(H)}. Pre-trained, foundational model using the transformer architecture & Single sequence (FASTA) & First to utilize a foundation model or pre-training approach & \href{https://github.com/ml4bio/RNA-FM}{Code}\\
    \end{longtable}
    \end{small}
\end{ThreePartTable}

\paragraph{\href{https://cs.au.dk/~compbio/pfold/downloads.html}{\textsc{Pfold} \& \textsc{PPfold}}}
\textsc{Pfold} is an RNA secondary structure predicting program that employs a stochastic context-free grammar ~\cite{knudsen1999rna, knudsen2003pfold}. 
As mentioned above, an SCFG is a probabilistic model that uses probabilities as parameters to measure the co-varying tendency of position pairs. 
The co-varying tendency assumes compensatory mutations at paired positions occur in a correlated way. 
Since the function of RNA sequences largely depends on their structures, evolutionarily related RNAs that exert similar functions are very likely to have similar structures. 
Thus, those highly co-varying positions across a set of evolutionarily related RNAs would maintain the structure (i.e. base pairing) even though the sequence similarity may be low. 
\textsc{Pfold} takes as input a multiple sequence alignment that contains target RNA homologous sequences in fasta format and predicts the consensus secondary structure of the alignment using the so called KH-99 algorithm, named after \textsc{Pfold}'s authors.
The KH-99 algorithm essentially couples a phylogenetic model calculated from the alignment using Felsenstain maximum likelihood algorithm~\cite{felsenstein1981evolutionary} with the SCFG, and it finds the most likely RSS using the CYK dynamic programming algorithm derived from natural language processing~\cite{sakai1961syntax}. 
An updated implementation of \textsc{Pfold} called \textsc{PPfold} was later developed\cite{sukosd2012ppfold} with Java multithreaded computation to accelerate the SCFG and phylogenetic calculations, which demonstrates much improved scalability. 
In addition to \textsc{Pfold}'s original output, \textsc{PPfold} generates a symmetric base-pairing probability matrix so that for each position, the probability of it being base-paired is computed as the average probability of all pairs that contain it.

\paragraph{\href{https://e-rna.org/rnadecoder/}{\textsc{RNAdecoder}}}
\textsc{RNAdecoder}~\cite{pedersen2004comparative, pedersen2004evolutionary} is a comparative method for making predictions of RNA secondary structure. It also employs an SCFG model but is more complex: it is explicitly designed to take protein-coding context into account. So for the input, besides the sequence alignment, the protein-coding annotation of the input mRNA alignment (specifying codon positions), as well as a phylogenetic tree for those sequences in the alignment, is also required. \textsc{RNAdecoder} is able to distinguish between loop/bulge region and un-paired region outside of the RNA structure and does not assume global RNA structure. \textsc{RNAdecoder} has two modes: (1) in scanning mode, it scans for posterior probabilities of a position being loop/bulge and unstructured in the alignment; (2) in folding mode, it predicts the RNA structure and explicitly labels stem-pairing, loop/bulge and unstructured positions. In contrast to the other programs, for a given position, \textsc{RNAdecoder} computes its probabilities for being loop/bulge and unstructured. So, the sum of these two probabilities is the non-base-pairing probability for that position.

\paragraph{\href{https://github.com/EddyRivasLab/tornado}{\textsc{TORNADO}}} \textsc{TORNADO}~\cite{rivas2012tornado} is a flexible SCFG-based approach and programming language itself designed for RSS prediction, which offers greater adaptability compared to the other SCFG method. It accepts RNA sequences in Stockholm or FASTA formats as input and outputs predicted secondary structures in dot-bracket notation. The method employs dynamic programming algorithms, such as CYK or posterior decoding, to predict structures and uses maximum likelihood optimization to train SCFG parameters from sequence-structure pairs. Its flexibility enables the modeling of complex structural elements, including base-pair stacking and loop dependencies, with SCFG models like the one used by \textsc{Pfold}~\cite{knudsen1999rna}. While this adaptability makes \textsc{TORNADO} capable of addressing a wide range of RNA modeling tasks, it still has a cubic time complexity \(O(KN^3)\) for a sequence of length $N$ and a design-dependent constant $K$, which can be a limitation for long sequences or grammars with high complexity.

\subsection{Advanced machine learning and deep learning based methods}
Since the classic methods such as the Zuker algorithm faced limitations due to the complexity and incompleteness of experimentally determined free energy parameters etc., machine learning and more recently deep learning methods have raised as powerful alternatives by leveraging large datasets of RNA sequences and their structures to provide pure data-driven, supervised-learning solutions.
These advanced learning-based methods have introduced sophisticated model parameterization and training frameworks, bypassing explicit thermodynamic or co-variation assumptions, and as a result, they significantly improved prediction accuracy in cases like pseudoknot and  long-range basepair interactions etc.

\paragraph{\href{http://contra.stanford.edu/contrafold/server.html}{\textsc{CONTRAfold}}} \textsc{CONTRAfold}~\cite{Do2006contrafold} is a machine learning-based method for RNA secondary structure prediction that employs a discriminative probabilistic model, conditional log-linear model (CLLM), instead of a SCFG-like generative model or relying solely on free energy minimization. 
It learns model parameters directly from the training data but resembles an MFE model for its CLLM construction.
Consequently, CONTRAfold aims to find the maximum-expected accuracy structure, which is the probabilistic counterpart of the MFE structure in energy-based models.
The method takes a single RNA sequence as input, and like many other methods, it uses a dynamic programming algorithm and has a time complexity $O(n^3)$.
As a result, the input RNA sequence is normally constrained under a practical upper limit (e.g., 1,000 as instructed on its web server) to maintain reasonable run times and memory usage. 
By incorporating benefits from both probabilistic and thermodynamic models, CONTRAfold demonstrates competitive or superior accuracy compared to purely SCFG-based methods such as Pfold and purely energy-based methods such as ViennaRNA.

\paragraph{\href{https://sparks-lab.org/server/spot-rna/}{\textsc{SPOT-RNA}}} \textsc{SPOT-RNA}~\cite{singh2019rna} is a deep learning-based RNA secondary structure prediction method that utilizes an ensemble of residual networks (ResNets)~\cite{he2016identity}, two-dimensional bidirectional long short-term memory (2D-BLSTM)~\cite{hochreiter1997long,schuster1997bidirectional} modules, and dilated convolutional neural networks (CNNs)~\cite{yu2015multi}.
It also leverages transfer learning to integrate information from a large dataset containing low-quality structure labels at a single base-pair level with a small but high-resolution training set.  
The model with transfer learning reduced the risk of overfitting and showed remarkable performance generalization across diverse datasets. 
Moreover, rather than relying on accurate thermodynamic parameters, SPOT-RNA adopts a pure machine-learning strategy so that all base pairs can be trained and predicted, regardless of whether it is associated with local or nonlocal (tertiary) interactions.
As a result, \textsc{SPOT-RNA} can predict base pairs beyond the canonical Watson–Crick interactions, including noncanonical and pseudoknotted (non-nested) configurations, which are often challenging for traditional prediction methods.

\paragraph{\href{https://sparks-lab.org/server/spot-rna2/}{\textsc{SPOT-RNA2}}} \textsc{SPOT-RNA2}~\cite{singh2021improved} is an extension builds upon \textsc{SPOT-RNA}~\cite{singh2019rna} by incorporating evolutionary information and refined neural architectures. 
Specifically, it uses an ensemble of dilated CNNs only to simplify the model architecture with faster computations for long-range interactions, and it reuses the transfer learning step to leverage the information in both the low-resolution and high-resolution structure data. 
In addition to the one-hot encoding and predicted base-pair probabilities from single sequences, SPOT-RNA2 integrates PSSMs (Position-Specific Scoring Matrices) and DCA (Direct Coupling Analysis) features, enabling it to capture evolutionary information as a supplement to improve prediction accuracy. 
By incorporating these heterogeneous inputs, SPOT-RNA2 outperforms the single-sequence-based SPOT-RNA, especially on highly homologous sequences, and enhances the ability to infer structurally complex patterns such as noncanonical pairs more effectively. 

\paragraph{\href{https://github.com/ml4bio/e2efold}{\textsc{E2Efold}}} \textsc{E2Efold}~\cite{Chen2020RNA} is an end-to-end deep learning model for RNA secondary structure prediction, designed to address challenges in dealing with complex RNA structures, particularly pseudoknots.
Unlike traditional approaches that rely on energy minimization and dynamic programming, it directly predicts the RNA base-pairing matrix while integrating hard constraints, through a deep architecture combining a transformer-based Deep Score Network for sequence representation and a multilayer Post-Processing Network for constraints on legit base-pairing types.
Specifically, the post-processing network employs an unrolled algorithm that utilizes a primal-dual constrained optimization to ensure the base-pairing constraints are enforced to reduce the space of valid structures and mitigate overfitting.
As a result, E2Efold achieved state-of-the-art performance at then, on benchmark datasets (we will describe in later section~\ref{sec: data}) including RNAStralign and ArchiveII, significantly improving prediction accuracy for both nested and pseudoknot structures.
A notable related work later introduced \textsc{E2Efold-3D} which is an extension of \textsc{E2Efold} for the de novo RNA tertiary structure prediction~\cite{shen2022e2efold3dendtoenddeeplearning}. 

\paragraph{\href{https://github.com/uci-cbcl/UFold?tab=readme-ov-file}
{\textsc{UFold}}}
\textsc{UFold}\cite{fu2022ufold} uses a deep CNN architecture called U-Net~\cite{ronneberger2015unet} to generate RNA secondary structure prediction. UFold provides both a web server and a local software option. It accepts multiple RNA sequences in FASTA format as input and outputs the predicted secondary structure in dot-bracket notation. Regarding its architecture, UFold converts the original sequence into an image of size $17 \times L \times L$, where $L$ is the length of the RNA sequence, using this image as the input to a U-Net architecture to generate a predicted contact map. The $17$ can be thought of as 17 different channels, 16 types of unique Waston-Crick base-pairing, and an extra channel used in CDPFold~\cite{hao2019cdpfold} to deal with sparsity. UFold's novelty came from converting raw RNA sequences into an ``image" representation, which allows for long-range contact predictions, a fully convolutional framework, and highly efficient parallel computability.

\paragraph{\href{https://github.com/mehdi1902/RNA-secondary-structure-prediction-using-CNN}
{\textsc{CNNFold}}} CNNFold\cite{saman2022rna} utilizes a simple yet effective CNN architecture. The method encodes RNA sequences of length $L$ into a two-dimensional $L \times L$ map with eight channels, where each channel represents specific base-pairing relationships or structural constraints. Six channels capture possible base pairings (e.g., A-U, G-C), one channel indicates unpaired bases along the main diagonal, and another flags invalid pairings due to constraints like short distances or incompatible bases. This representation facilitates the prediction of both local and long-distance pairings, as well as structural motifs like stems. The output of the model is an $L \times L$ base-pair scoring matrix. To ensure the validity of predicted structures, CNNFold modifies the Blossom algorithm to handle self-loops, which is crucial for representing and predicting structures like pseudoknots. While CNNFold doesn't particularly stand out in complexity, it serves as a notable example of a biologically informed architecture and its distinct post-processing techniques.

\paragraph{\href{https://redfold.ee.ncyu.edu.tw/redfold/}
{\textsc{REDfold}}}
\textsc{REDfold}\cite{chen2023redfold} is another deep learning-based model that is unique in using an encoder-decoder network incorporating ResNet~\cite{he2016deep} and FC-DenseNet~\cite{jegou2017one} networks. It provides both a web server and the source code for the software. As input, it takes FASTA formatted RNA sequence files, and for output, it outputs in the dot-bracket notation. REDfold, much like UFold, first generates the RNA sequence into two-dimensional contact matrices by trying all possible combinations of Waston-Crick base pairing. These matrices are fed into the network via feature mapping and basic convolution modules (BCMs) consisting of 2-dimensional convolution, batch-normalization, and rectified linear unit (ReLU). They also introduce a dense connected module (DCM), which is made up of a series of BCMs to avoid bottlenecks from the encoding steps. The decoder network also consists of DCMs that ultimately transition up to generate a scoring matrix of size $L \times L$ where $L$ is the length of the RNA sequence. REDfold claims that the incorporation of ResNet and FC-DenseNet networks in their encoder-decoder networks makes the process much more efficient and effective, producing highly accurate predictions.

\subsection{Hybrid methods}
 
Despite the recent advances in the class of learning-based methods, its reliance on large and unbiased datasets highlights challenges such as overfitting and generalization to complex structural patterns. 
To overcome these challenges, efforts from the wet-lab have been dedicated to improving both the quantity and quality of RNA structural data, while computationally, exploring less data-hungry approaches, such as hybrid methods that integrate constraints to the deep model space as well as transfer learning from pre-trained general foundation models, seems to offer a promising path.

\paragraph{\href{rna.tbi.univie.ac.at/cgi-bin/RNAalifold.cgi}{\textsc{RNAalifold}}}
\textsc{RNAalifold}~\cite{bernhart2008rnaalifold} is a program in the ViennaRNA package, representing an early work adopting hybrid strategy to combine energy and comparative approaches. 
As the name suggests, it computes the minimum energy structure for a set of \textit{aligned} input sequences. 
In contrast to \textsc{Pfold}, \textsc{RNAalifold} uses free energies rather than probabilities as parameters.
It combines the co-varying information from the fixed alignment with the minimum free energy model by modifying the scoring scheme of the dynamic programming algorithm used in conventional thermodynamic methods.
It produces the consensus minimum free energy structure in dot-bracket notation and a dot plot of the symmetric base-pairing probability matrix. 

\paragraph{\href{http://rtools.cbrc.jp/centroidfold/}{\textsc{CentroidFold}}} \textsc{CentroidFold} is based on the $\gamma$-centroid estimator \cite{hamada2009prediction} for high-dimensional discrete spaces, which is generally more accurate than an maximum expected accuracy estimator (e.g. the one used in CONTRAfold~\cite{Do2006contrafold}) under the same probability distribution. \textsc{CentroidFold} supports multiple probabilistic models, including the CONTRAfold model, the McCaskill model (from the ViennaRNA package), the RNAalifold model, and the Pfold model. Benchmarks indicate that CentroidFold with the McCaskill model using Boltzmann likelihood parameters \cite{andronescu2010computational} achieves the most accurate predictions. \textsc{CentroidFold} accepts RNA sequences in FASTA format as input, producing the predicted secondary structure.

\paragraph{\href{https://ws.sato-lab.org/mxfold/}{\textsc{MxFold}}}
In order to deal with overfitting issues in learning-based RSS prediction methods, Akiyama et al. introduced a novel method \textsc{MxFold}, which integrates thermodynamic information and machine learning models together. 
It accepts RNA sequences in FASTA or bpseq format as input and outputs predicted secondary structures through a Zuker-style dynamic programming (DP) algorithm~\cite{zuker1981optimal}. 
The DP algorithm predicts an optimal secondary structure that maximizes the sum of scores from a hybrid model.
The model combines the Turner's nearest-neighbor parameters~\cite{mathews2004incorporating, mathews288expanded}, which are experimentally determined free-energy parameters, with a fine-grained scoring model based on structured support vector machines (SSVMs). 
This hybrid model trains weights of SSVMs for complex structural features, such as specific loop configurations and base-pair stacking. 
By combining thermodynamic parameters with machine-learned scores, the method ensures both accuracy and robustness, particularly for unobserved substructures. 
For example, given a sequence, \textsc{MxFold} decodes the structure using a scoring function:

\begin{equation}
f(x, y) = f_T(x, y) + f_W(x, y)
\label{eqn:mxfold}
\end{equation}

where \(f_T\) encapsulates thermodynamic contributions, and \(f_W\) reflects the machine learning-based refinements. 
This hybrid framework demonstrates superior prediction performance while mitigating risks of overfitting, thereby advancing RNA secondary structure prediction.

\paragraph{\href{https://ws.sato-lab.org/mxfold2/}{\textsc{MxFold2}}}
Inspired by the success of deep learning in biological sequence analysis,  Sato et al. further extended \textsc{MxFold}~\cite{akiyama2017mxfold} with deep neural networks replacing the SSVMs in the model part and developed \textsc{MxFold2}~\cite{sato2021rna}. 
\textsc{MxFold2} adopts a combination of CNN and BiLSTM layers for the network architecture to learn four different types of folding parameters: helix stacking, helix opening, helix closing, and unaired region scores. 
These parameters are then fed into a Zuker-style dynamic programming algorithm~\cite{zuker1981optimal}, just like the Turner nearest neighbor parameters, to calculate the final score of an RNA secondary structure.
Similar to \textsc{MxFold}, \textsc{MxFold2} integrates the thermodynamic information to reduce overfitting by separately computing minimum free energy scores through the same DP algorithm as well and added to the deep neural network scores as a form of regularization.
As a result, MXfold2 aligns deep learning folding scores with free energy folding scores and achieves superior performance compared to several traditional and DNN-based models, including MXfold, across sequence-wise and family-wise testing datasets.

\paragraph{\href{https://github.com/ml4bio/RNA-FM}
{\textsc{RNA-FM}}} RNA-FM\cite{chen2022interpretable} represents one of the first approaches to integrate pre-training. Unlike previous deep learning-based strategies, which rely on labeled data specific to secondary structure, RNA-FM leverages the vast pool of unannotated RNA sequence data through self-supervised learning. Its architecture is based on a transformer model comprising 12 bidirectional encoder layers, pre-trained on 23 million RNA sequences from RNAcentral by reconstructing masked tokens. This approach enables RNA-FM to learn rich, task-agnostic representations of RNA sequences, capturing implicit structural and evolutionary patterns without requiring labels. These embeddings are then fine-tuned for downstream structure-related and function-related applications, offering flexibility across diverse RNA prediction tasks. RNA-FM’s advantages are more pronounced in structural tasks, like secondary structure prediction, likely due to differences between its training data and the datasets used for functional tasks, as well as the inherent complexity of RNA structure-function relationships.

\paragraph{\href{https://github.com/CatIIIIIIII/RNAErnie}
{\textsc{RNAErnie}}} RNAErnie\cite{wang2024multi} is a recent foundational model, also pre-trained on RNAcentral data, that distinguishes itself through the integration of motif-aware pretraining strategies and a type-guided fine-tuning mechanism. RNAErnie features 12 multilayer transformer blocks with a hidden state dimension of 768. While RNA-FM had base-level masking alone, RNAErnie's pretraining phase incorporates a motif-aware multilevel masking strategy, which includes base-level, subsequence-level, and motif-level masking. This approach enriches RNA representations by capturing both fundamental sequence patterns and biologically significant motifs, such as those derived from databases like ATtRACT and SpliceAid. Additionally, RNAErnie tokenizes coarse-grained RNA types (e.g., miRNA, lnRNA) as special vocabularies, appending them to RNA sequences during pretraining to improve domain adaptation and enhance the model’s understanding of RNA-specific features.

For the downstream task, RNAErnie employs a type-guided fine-tuning strategy, which predicts RNA types from sequence embeddings and incorporates these as auxiliary inputs into task-specific modules. Their study examines multiple architectures: FBTH (frozen backbone with trainable task-specific head), TBTH (trainable backbone and head for end-to-end learning), and, novelly, STACK (ensemble learning with type-guided parallel modules). For secondary structure prediction, RNAErnie combines its embeddings with a  Zuker-style dynamic programming approach like \textsc{MXFold}, predicting RNA secondary structure by maximizing the cumulative scores of adjacent loops. Fine-tuning is performed using a max-margin framework, minimizing structured hinge loss with thermodynamic regularization.

\section{Computational tools for RNA modification prediction}
This section reviews recent tools for several common RNA modification types, including 2'-O-methylation (Nm or 2OM), N4-acetylcytosine (ac4C), 5-methylcytosine (m5C), N6-methyladenosin (m6A), N7-methylguanosine (m7G), and multiple widely occurring transcriptome modifications. Table~\ref{RM_tools} summarizes these tools, with detailed descriptions provided below.

\begin{ThreePartTable}
    \begin{TableNotes}\footnotesize
        \item \textbf{Note:} Bi-GRU: Bidirectional gated recurrent unit. Bi-LSTM: Bidirectional long short-term memory. CNN: Convolutional neural networks. nt: nucleotides. These tools are grouped based on the method category.
    \end{TableNotes}
    \begin{small}    
    \begin{longtable}{lp{0.23\textwidth}p{0.17\textwidth}p{0.23\textwidth}p{0.09\textwidth}}
    \caption{A summary of representative RNA modification prediction methods included in the review.}
    \label{RM_tools} \\
    \hline
    \textbf{} & \textbf{Method} & \textbf{Input} & \textbf{Other notes} & \textbf{URLs} \\
    \hline
    \endfirsthead
    \multicolumn{5}{l}{Continuation of table}\\
    \hline
    \textbf{} & \textbf{Method} & \textbf{Input} & \textbf{Other notes} & \textbf{URLs} \\
    \hline
    \endhead
    \hline
    \multicolumn{5}{r}{More on next page}\\
    \endfoot
    \hline
    \insertTableNotes
    \endlastfoot
    \textbf{{\sc H2Opred}}~\cite{pham2024h2opred} & Hybrid deep learning with multi-feature fusion & Multiple sequences or file in FASTA format & No more than 500 sequences per submission & \href{https://balalab-skku.org/H2Opred/}{Web server} \\
    \textbf{{\sc Meta-2OM}}~\cite{harun2024meta} & Multi-classifier meta-learning & Multiple sequences or file in FASTA format &  & \href{https://github.com/kuratahiroyuki/Meta-2OM/}{Code}, \href{http://kurata35.bio.kyutech.ac.jp/Meta-2OM/}{Web server} \\
    \textbf{{\sc Nmix}}~\cite{geng2024nmix} & Hybrid deep learning with multi-feature fusion and ensemble learning & Multiple sequences or file in FASTA format & Up to 5000 sequences per submission & \href{https://github.com/tubic/Nmix/}{Code}, \href{http://origin.tubic.org/Nm/public/index.php/index}{Web server} \\
    \textbf{{\sc ac4C-AFL}}~\cite{pham2024ac4c} & Adaptive feature representation learning & Multiple sequences or file in FASTA format & No more than 20 sequences per submission & \href{https://balalab-skku.org/ac4C-AFL/}{Web server} \\
    \textbf{{\sc Voting-ac4C}}~\cite{jia2024voting} & Pre-trained large RNA language model and ensemble learning & Multiple sequences or file in FASTA format &  & \href{http://www.bioai-lab.com/ac4C/}{Web server} \\
    \textbf{{\sc iRNA-ac4C}}~\cite{su2023irna} & Machine learning, minimum-Redundancy-Maximum-Relevance combined with incremental feature selection strategies & Multiple sequences in FASTA format & Sequence length must be 201 nt & \href{http://lin-group.cn/server/iRNA-ac4C/}{Web server} \\
    \textbf{{\sc TransAC4C}}~\cite{liu2024transac4c} & Transformer-based encoder and Bi-LSTM networks combined with 1D CNN & Multiple sequences or file in FASTA format & Select either 415 nt or 21 nt & \href{https://github.com/BioLJ/TransAC4C/}{Code} \\
    \textbf{{\sc Deepm5C}}~\cite{hasan2022deepm5c} & Hybrid deep learning & CSV format file &  & \href{https://github.com/hasan022/Deepm5C/}{Code} \\
    \textbf{{\sc MLm5C}}~\cite{kurata2024mlm5c} & A combination of hybrid machine learning models & Multiple sequences or file in FASTA format &  & \href{https://github.com/kuratahiroyuki/MLm5C/}{Code}, \href{http://kurata35.bio.kyutech.ac.jp/MLm5C/}{Web server} \\
    \textbf{{\sc m5C-pred}}~\cite{abbas2023xgboost} & XGBoost framework with feature selection & Multiple sequences in FASTA format & Select the species for prediction & \href{https://github.com/Z-Abbas/m5C-pred/}{Code}, \href{https://nsclbio.jbnu.ac.kr/tools/m5C-pred/}{Web server} \\
    \textbf{{\sc MST-m6A}}~\cite{su2024mst} & Multi-scale transformer-based framework & TSV format file & Sequence length must be 201 nt & \href{https://github.com/cbbl-skku-org/MST-m6A/}{Code} \\
    \textbf{{\sc CLSM6A}}~\cite{zhang2023interpretable} & Interpretable deep learning-based approach & Multiple sequences or file in FASTA format & Sequence length must be 201 nt & \href{https://github.com/zhangying-njust/CLSM6A/}{Code}, \href{https://csbioinformatics.njust.edu.cn/clsm6a/}{Web server} \\
    \textbf{{\sc BLAM6A-Merge}}~\cite{xia2024blam6a} & Attention mechanisms with multimodal feature fusion and Blastn tool & FASTA format file & Required Blastn of version 2.14.0 & \href{https://github.com/DoraemonXia/BLAM6A-Merge/}{Code} \\
    \textbf{{\sc Moss-m7G}}~\cite{zhao2024moss} & Motif-based interpretable deep learning & FASTA format file &  & \href{https://github.com/MrQcx/Moss-m7G/}{Code} \\
    \textbf{{\sc THRONE}}~\cite{shoombuatong2022throne} & Three-layer ensemble learning & Multiple sequences or file in FASTA format &  & \href{http://thegleelab.org/THRONE/}{Web server} \\
    \textbf{{\sc MultiRM}}~\cite{song2021attention} & Attention-based multi-label neural networks  & Single RNA sequence in string format & Minimum length of 51 nt & \href{https://github.com/Tsedao/MultiRM/}{Code} \\
    \textbf{{\sc TransRNAm}}~\cite{chen2023transrnam} & Transformer-based encoder and CNN &  &  & \href{https://github.com/lennylv/TransRNAm/}{Code} \\
    \textbf{{\sc CIL-RNA}}~\cite{qiao2024towards} & Transformer-based encoder and Bi-GRU network with class incremental learning & CSV format file &  & \href{https://github.com/KazeDog/cl_rna/}{Code} \\
    \hline
    \end{longtable}
    \end{small}
\end{ThreePartTable}

\subsection{2'-O-methylation (Nm or 2OM)}
\paragraph{\href{https://balalab-skku.org/H2Opred/}{\textsc{H2Opred}}}
\textsc{H2Opred} represents the first hybrid deep learning framework developed for the identification of 2OM in human RNA. This framework utilizes a combination of stacked one-dimensional convolutional neural networks (1D CNN) and attention-based bidirectional gated recurrent unit (BiGRU) modules to effectively capture both spatial and temporal information derived from conventional feature descriptors and embeddings based on natural language processing (NLP). The resultant high-level feature representations are subsequently integrated to facilitate the final classification of nucleotide modifications, specifically Am, Cm, Gm, Um, or Nm. \textsc{H2Opred} thus accommodates both nucleotide-specific and generic 2OM modification types. The associated web server accepts input in FASTA format, providing users with probabilistic scores and class labels for the corresponding sequences. Additionally, users have the option to upload a FASTA file to the web server to execute predictions and retrieve results.

\paragraph{\href{http://kurata35.bio.kyutech.ac.jp/Meta-2OM/}{\textsc{Meta-2OM}}}
\textsc{Meta-2OM} is a multi-classifier meta-learning approach that integrates eight distinct machine learning classifiers with eighteen different feature encoding algorithms, all coordinated by a meta-learner, to identify 2OM in human RNA. Notably, probabilistic features from 144 baseline models were generated and subsequently utilized to train a logistic regression model for the final classification task. This tool is available as both a web server and a downloadable codebase, accommodating the analysis of multiple sequences or file uploads in FASTA format. Upon completion of the submission process, the system returns probabilistic scores and class labels corresponding to the input sequences.

\paragraph{\href{http://origin.tubic.org/Nm/public/index.php}{\textsc{Nmix}}}
\textsc{Nmix} is a hybrid deep learning framework developed for the identification of 2OM sites in human RNA. Initially, one-hot, Z-curve, and RSS encodings were extracted from the RNA sequences. Subsequently, 1D and 2D CNN were designed, incorporating multi-head self-attention and residual connection modules to extract multi-dimensional features from the one-hot and Z-curve encodings as well as the RSS encoding. These feature representations were later fused through average pooling and concatenation for the purpose of final classification. Additionally, a Bayesian optimization-based technique was employed to construct an ensemble learning framework that effectively addresses the challenges presented by imbalanced datasets. Given an RNA sequence, \textsc{Nmix} outputs the nucleotide base, a probabilistic score, and a corresponding class label.

\subsection{N4-acetylcytosine (ac4C)}
\paragraph{\href{https://balalab-skku.org/ac4C-AFL/}{\textsc{ac4C-AFL}}}
\textsc{ac4C-AFL} is an adaptive feature representation learning framework designed for the identification of ac4C in human mRNA. Initially, a pre-analysis was conducted to determine the optimal sequence length for ac4C identification, leading to the conclusion that a length of 201 nucleotides (nt) is optimal. Subsequently, a novel ensemble feature importance scoring function was proposed to identify the optimal feature dimensions from sixteen sequence-derived feature descriptors, employing a sequence forward search strategy. Utilizing these optimal features, 176 single-feature best-performing models were constructed using eleven distinct machine learning algorithms, and their probabilistic features were generated to train the final model for ac4C identification. \textsc{ac4C-AFL} is publicly accessible, allowing users to input sequences in FASTA format or directly upload FASTA files to obtain predicted results, including probabilistic scores and class labels.

\paragraph{\href{http://www.bioai-lab.com/ac4C}{\textsc{Voting-ac4C}}}
\textsc{Voting-ac4C} is the first framework that harnesses the capabilities of the pre-trained large RNA language model, RNAErnie~\cite{wang2024multi}, in conjunction with six conventional feature descriptors: one-hot encoding, encoding nucleic acid composition (ENAC), C2, nucleotide density (ND), trinucleotide composition profile (TPCP), and k-spaced nucleotide pair frequencies (KSNPF). This integration aims to enhance the prediction of RNA ac4C sites. A deep neural network (DNN) model was specifically designed for feature reduction and selection. Subsequently, a soft voting ensemble learning model was constructed by integrating eXtreme Gradient Boosting (XGB), CatBoost (CB), and multilayer perceptron (MLP) for the final prediction. Similar to other tools, \textsc{Voting-ac4C} accepts multiple sequences in FASTA format for predicting ac4C sites.

\paragraph{\href{https://github.com/BioLJ/TransAC4C/}{\textsc{TransAC4C}}}
\textsc{TransAC4C} is an interpretable framework that employs a transformer-based encoder to leverage the relationships between words in natural language sequences, translating these relationships into biological contexts for model interpretation. Notably, this study involved reconstructing a previous dataset to generate a new dataset characterized by varying sequence lengths, distinct species, and diverse RNA types, thereby facilitating a comprehensive analysis of ac4C in RNA. Specifically, RNA sequences were tokenized using 3-mers and subsequently embedded as inputs for a transformer-based bidirectional long short-term memory (Bi-LSTM) module, which extracts contextual information. A 1D convolutional neural network (CNN) module was then designed to capture essential spatial information before processing the features through several fully connected (FC) layers for final classification. This tool is capable of predicting ac4C from multiple RNA sequences in FASTA format. However, it requires the selection of either 415 or 21 nt corresponding to the appropriate models.

\paragraph{\href{http://lin-group.cn/server/iRNA-ac4C/}{\textsc{iRNA-ac4C}}}
\textsc{iRNA-ac4C} is a machine learning-based predictor designed for the identification of ac4C in mRNA. A novel high-quality dataset was constructed to develop this ac4C prediction tool, utilizing the Gradient Boosting Decision Tree (GBDT) classifier along with optimal hybrid features. These optimal features were identified by linearly combining k-mer encoding, nucleotide chemical property encoding, and accumulated nucleotide frequency encoding, followed by the application of minimum redundancy maximum correlation (mRMR) and incremental feature selection (IFS) techniques to select the optimal feature dimensions. \textsc{iRNA-ac4C} offers a publicly accessible web server that supports the input of multiple sequences in FASTA format, specifically with a fixed length of 201 nt.

\subsection{5-methylcytosine (m5C)}
\paragraph{\href{https://github.com/hasan022/Deepm5C/}{\textsc{Deepm5C}}}
\textsc{Deepm5C} is a deep learning (DL)-based hybrid stacking tool developed for the prediction of m5C in the human genome. Initially, a novel benchmark dataset was constructed, and four distinct feature encodings were employed to extract relevant features, which included three conventional feature descriptors and a natural language processing (NLP)-based embedding known as word2vec. Subsequently, four DL-based classifiers and four machine learning (ML)-based classifiers were utilized to train a total of 32 baseline models. The probabilistic features generated from these models were then stacked to train the final model using a one-dimensional convolutional neural network (1D CNN). While this tool provides the source code, the instructions are brief. It supports input files in CSV format.

\paragraph{\href{http://kurata35.bio.kyutech.ac.jp/MLm5C/}{\textsc{MLm5C}}}
\textsc{MLm5C} a hybrid ML-based model designed to identify m5C sites. This model combines four ML classifiers with eleven RNA sequence-derived conventional feature descriptors. Subsequently, 44 single-feature baseline models were generated and ranked based on their performance, with the probabilistic features of the top 20 models stacked to train the final predictive model for m5C identification. Although this tool demonstrates superior performance compared to state-of-the-art methods at the time, its approach resembles several publications from the same author group. \textsc{MLm5C} framework provides both a web server and source code for making predictions and facilitating local deployment. It supports the analysis of multiple sequences or the upload of files in FASTA format. Upon completing the submission job, the tool returns probabilistic scores and class labels for the input sequences.

\paragraph{\href{https://nsclbio.jbnu.ac.kr/tools/m5C-pred/}{\textsc{m5C-pred}}}
\textsc{m5C-pred} is an ML-based framework that incorporates a feature selection strategy to predict m5C sites in five different species: Arabidopsis thaliana, Danio rerio, Drosophila melanogaster, Homo sapiens, and Mus musculus. This tool employs the XGB algorithm and utilizes five conventional feature descriptors, including the composition of k-spaced nucleic acid pairs (CKSNAP), enhanced nucleic acid composition (ENAC), label encoding (LE), nucleotide chemical properties (NCP), and electron-ion interaction pseudopotentials of trinucleotides (PseEIIP). Additionally, SHapley Additive exPlanations (SHAP) analysis is employed to identify the optimal features, which are then used to retrain the XGB for the final model. \textsc{m5C-pred} framework offers both a web server and source code; however, it necessitates the selection of a specific species for making predictions.

\subsection{N6-methyladenosin (m6A)}
\paragraph{\href{https://csbioinformatics.njust.edu.cn/clsm6a/}{\textsc{CLSM6A}}}
\textsc{CLSM6A} is an interpretable DL-based architecture developed for the prediction of N6-methyladenosine (m6A) modification sites across various cell lines and tissues in Homo sapiens. Specifically, RNA sequences are transformed into a 2D matrix using an ENAC-based encoding algorithm. A CNN module is then employed to extract and learn the spatial information from these features, which are subsequently input into an MLP for final classification. Additionally, both model-based and propagation-based methods are utilized to interpret the predictions made by the model. Furthermore, \textsc{CLSM6A} offers both a web server and publicly available source code for prediction and local deployment. However, the input sequence length must be precisely 201 nt.

\paragraph{\href{https://github.com/cbbl-skku-org/MST-m6A/}{\textsc{MST-m6A}}}
\textsc{MST-m6A} is a multi-scale dual transformer-based architecture designed for the accurate identification of m6A modification sites across eight cell lines and three tissues in Homo sapiens. This framework employs a shared transformer architecture coupled with dual k-mer tokenization to exploit multi-scale feature representations from RNA sequences, thereby capturing global contextual information and enriching feature representations. These feature representations are subsequently fused using a channel feature fusion module and processed through three CNN layers before being input into an MLP module for final classification. Additionally, this tool provides publicly available source code to facilitate local deployment and ensure reproducibility of results.

\paragraph{\href{https://github.com/DoraemonXia/BLAM6A-Merge/}{\textsc{BLAM6A-Merge}}}
\textsc{BLAM6A-Merge} is a tool designed for the identification of m6A modification sites across twelve benchmark datasets derived from six cell lines, including CD8T, Hek293\_abacm, Hek293\_sysy, HeLa, and MOLM13, and operates in two modes: full transcript and mature mRNA. Notably, \textsc{BLAM6A-Merge} employs various attention-based mechanisms to extract multimodal features from RNA sequences. Subsequently, a stacking ensemble learning framework is utilized to integrate four specific classifiers derived from sequence data with a Blastn-based classifier, establishing a meta-learning approach for final classification. This tool provides source code for local deployment, facilitating the reproducibility of results. However, it is essential to note that using Blastn tool version 2.14.0 is required to generate the final predictions.

\subsection{N7-methylguanosine (m7G)}
\paragraph{\href{https://github.com/MrQcx/Moss-m7G/}{\textsc{Moss-m7G}}}
\textsc{Moss-m7G} is an interpretable DL-based method designed for the identification of m7G sites, utilizing word-detect and motif-based embedding within a transformer architecture. The word-detect module employs a 1D CNN to capture the motif probabilistic matrix derived from the one-hot encoding of RNA sequences. This motif probabilistic matrix is then transformed into motif-based embeddings and combined with an additional [CLS] token embedding, serving as input for the transformer architecture to capture high-level contextual information. Finally, the feature representation extracted from the transformer architecture through the [CLS] token is fed into an MLP module for final classification. This tool provides source code for making predictions and accepting input in FASTA format files.

\paragraph{\href{http://thegleelab.org/THRONE/}{\textsc{THRONE}}}
\textsc{THRONE} is a three-step ensemble learning framework developed for the identification of m7G sites in human RNA. Initially, nine conventional feature descriptors and six machine learning (ML)-based classifiers were utilized to generate 54 single-feature ML-based baseline models. Subsequently, the probabilistic features extracted from these baseline models were concatenated into a 54-dimensional feature vector and trained using six ML-based classifiers. Finally, the probabilistic features from the six ML-based meta-models served as input for a six-dimensional feature vector, which was used to identify the best meta-learner model for final classification. Furthermore, \textsc{THRONE} offers a web server that facilitates predictions of m7G sites by allowing users to upload files or directly enter multiple sequences in FASTA format.

\subsection{Multiple widely occurring transcriptome modifications}
\paragraph{\href{https://github.com/Tsedao/MultiRM/}{\textsc{MultiRM}}}
\textsc{MultiRM} is an attention-based DL framework designed for the prediction of twelve prevalent transcriptome modifications, including m1A, m5C, m5U, m6A, m6Am, m7G, $\Psi$, A-to-I, and four types of 2OM modifications. Initially, one-hot encoding is employed to convert the RNA sequence into a matrix format. Subsequently, three embedding strategies are applied to extract features from the one-hot encoded input, specifically utilizing 1D CNN, Word2Vec, and Hidden Markov Models. These feature representations are then fused and processed through an attention-based LSTM network for multi-label classification to predict the twelve RNA modification types. Moreover, the attention weights and integrated gradients are utilized to identify sequence motifs corresponding to those identified by motif sequence-based tools for each RNA modification. Although \textsc{MultiRM} offers both a web server and source code for predicting RNA modifications and facilitating local deployment, it is important to note that the web server is inactive at the time of this review.

\paragraph{\href{https://github.com/lennylv/TransRNAm/}{\textsc{TransRNAm}}}
\textsc{TransRNAm} is an interpretable transformer-based architecture designed for the identification of twelve common RNA modifications. Initially, Word2Vec is employed to convert RNA sequences into matrix representations. Subsequently, a transformer-based encoder is utilized to learn high-level contextual information from the features extracted via Word2Vec. These feature representations are then processed through a CNN block with a skip connection to capture spatial feature representations effectively. Finally, these spatial features are fed into twelve parallel FC networks to predict the twelve types of RNA modifications simultaneously. Furthermore, the attention weights of \textsc{TransRNAm} are extracted for model interpretation purposes. Although \textsc{TransRNAm} provides its implementation through a GitHub repository, it is noteworthy that there is no associated web server for online predictions.

\paragraph{\href{https://github.com/KazeDog/cl_rna/}{\textsc{CIL-RNA}}}
\textsc{CIL-RNA} is a class incremental learning framework designed to predict multiple types of RNA modifications. This framework employs a baseline classifier integrated with a transformer-based encoder, followed by a Bi-GRU and an MLP for final predictions. Specifically, it utilizes four incremental learning strategies to train the baseline classifier: parameter regularization, function regularization, replay, and template-based classification. Notably, \textsc{CIL-RNA} can be extended to predict new categories of RNA modification sites without retraining on previous data, thereby enhancing computational efficiency. To facilitate local deployment and ensure reproducibility, \textsc{CIL-RNA} provides its implementation via a GitHub repository and supports input files in CSV format.

\section{The interplay between RNA secondary structure and RNA modification} 
\label{sec:application}

\begin{figure}[!tpb]
  \begin{center}
    \includegraphics[width = \textwidth]{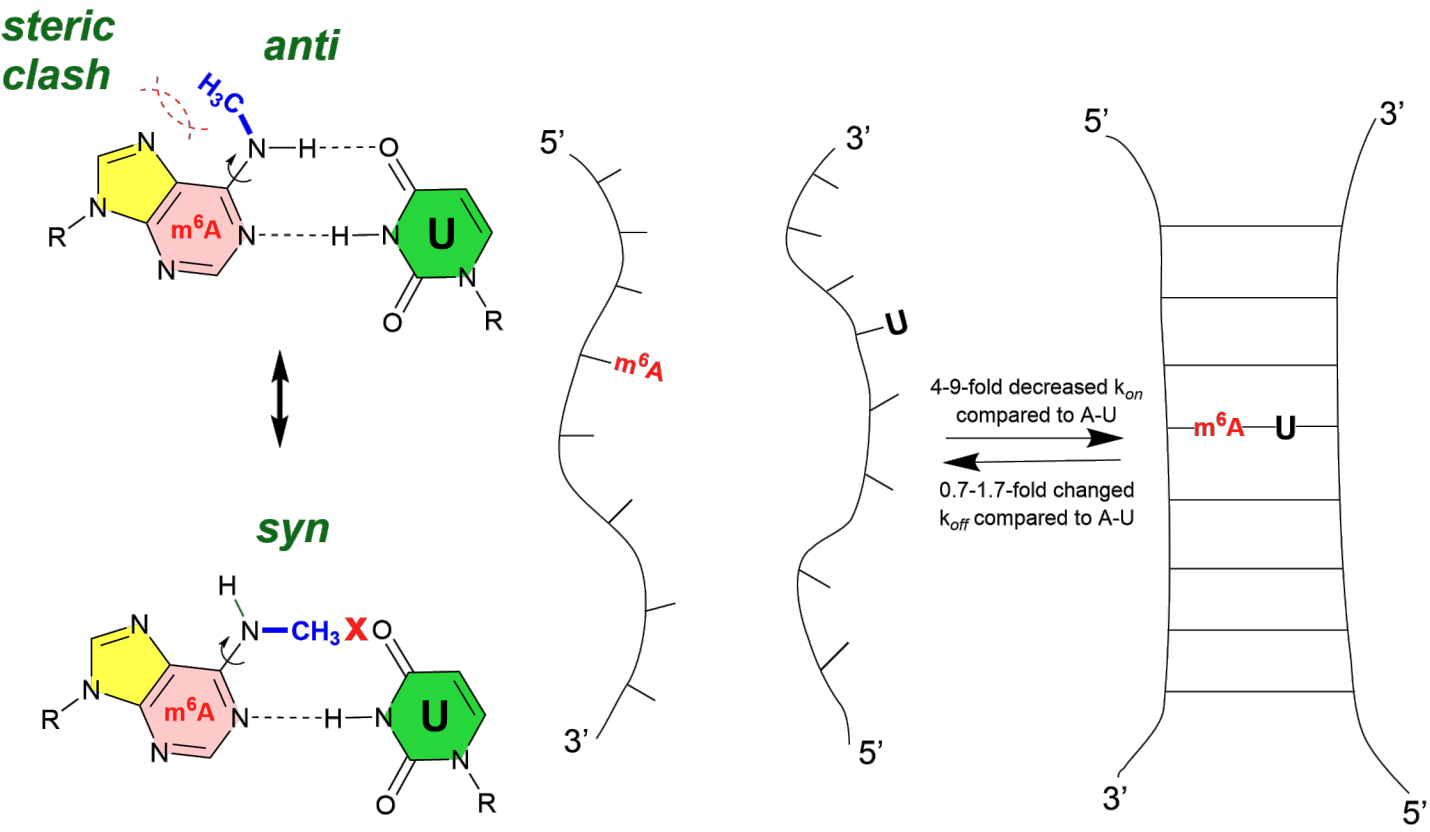}
    \caption{\textbf{An example of the interplay between RNA modification and secondary structures.} The m6A modification is used here to illustrate the effects of modifications on RNA secondary structures. The left plot shows that stable m6A:U basepairing is only feasible in the anti-conformation of the m6A base, which is energetically less favorable. The right plot shows the m6A:U base-pairing has a significant change on the annealing rate constant while the unpairing does not, when comparing to the normal A:U case. The figure is remade based on Hofler and Duss (2024)~\cite{hofler2024interconnections}. 
    }\label{modification} 
  \end{center}
\end{figure}

\subsection{RSS motifs aid RNA modifications prediction}

RNA structure plays a crucial role in RNA modifications, particularly in the m6A modification~\cite{hofler2024interconnections}. The folding and structural characteristics of RNA influence not only how and where these modifications are applied but also their interactions with proteins. Such RNA binding proteins reply on specific RSS patterns to bind to target RNA and initiate RNA modifications. For example, the m6A writer complex (METTL3/METTL14) is recruited to RNA based on these structural features, enabling site-specific and co-transcriptional installation of the m6A modification. This recruitment is facilitated by interactions with histone marks and the C-terminus of RNA Polymerase II, ensuring that the modification occurs precisely at the required locations during transcription. Specifically, the METTL3/METTL14 complex preferentially recognizes and modifies RNA sequences that contain specific motifs, particularly the DRACH motif (D = A/G/U; R = A/G; H = A/C/U). This sequence specificity ensures that the complex accurately targets the most relevant RNA substrates for modification.

Recently, researchers have integrated RNA secondary structure-based feature encodings to improve the performance of RNA modifications~\cite{el2021machine}. For instance, Xiang~\textit{et~al.}~\cite{xiang2016rnamethpre} utilized RNAfold (reviewed above) to fold the 101-bp mRNA fragment, yielding an MFE (minimum free energy) value. Then, it was combined with conventional feature descriptors to train a support vector machine (SVM) classifier to identify mRNA m6A sites. Similarly, Geng~\textit{et~al.}~\cite{geng2024nmix} also utilized RNAfold to generate structural expressions for each sequence. Using an ensemble learning approach, they combined it with one-hot encoding and Z-curve theory to enhance the prediction of 2’-O-methylation sites. These findings suggest that RNA secondary structure is crucial for understanding RNA’s biological functions and properties, leading to improved performance of RNA modifications.

\subsection{RNA modifications affect secondary structure formation}

RNA modifications, such as N6-methyladenosine (m6A), have a profound impact on RNA secondary structure prediction by altering the chemical and physical properties of RNA molecules~\cite{hofler2024interconnections}. Methylation of nucleotides modifies their characteristics, influencing RNA structure and interactions with cellular partners. These modifications can either promote or hinder the formation and functionality of protein-RNA complexes, as well as alter the base-pairing kinetics of RNA. For instance, the addition of a methyl group to the nitrogen position of adenosine affects the stability of base pairs, consequently slowing the rate of duplex formation. This alteration may lead to local destabilization of RNA structures, which standard secondary structure prediction algorithms often overlook.

In the study conducted by Liu~\textit{et~al.}~\cite{liu2015n}, it was demonstrated that m6A plays a crucial role in regulating RNA-protein interactions by modulating the accessibility of RNA-binding motifs (RBMs). This modification can reshape the local structure of mRNA and long non-coding RNA (lncRNA), thereby facilitating the binding of RNA-binding proteins such as heterogeneous nuclear ribonucleoprotein C (HNRNPC). The presence of m6A enhances the binding affinity of these proteins, influencing essential processes such as pre-mRNA processing, gene expression, and RNA maturation. This phenomenon, often referred to as the “m6A-switch,” describes how m6A-dependent structural remodeling of RNA regulates interactions between RNA and proteins, allowing for effective access to binding sites that are vital for various biological functions.

Moreover, Lewi~\textit{et~al.} provided evidence supporting the notion that both RNA structures and RNA modifications collaboratively shape RNA–protein interactions~\cite{lewis2017rna}. Specifically, m6A has been shown to destabilize stem structures, which enhances the accessibility of RNA-binding proteins to their binding sites. This dynamic interplay between RNA modifications and structure is crucial for regulating various aspects of gene expression, including mRNA stability, splicing, and translation efficiency. The structural changes induced by RNA modifications not only facilitate the recruitment of specific RNA-binding proteins (RBPs) but also influence the fate of mRNAs, such as their translation or decay, contributing to the overall regulation of gene expression.

Additionally, Tanzer~\textit{et~al.} asserted that RNA modifications can significantly impact RNA structures by either stabilizing or destabilizing base pairs~\cite{tanzer2019rna}. For example, A-to-I editing can destabilize double-stranded RNAs (dsRNAs) by converting A-U pairs into less stable I-U pairs, which increases flexibility and potential for refolding. Conversely, this editing can create stabilizing I-C pairs that enhance hybridization stability in certain contexts. Furthermore, m6A modification is known to weaken RNA structures and is particularly abundant in 3’ UTRs, influencing mRNA preprocessing events such as splicing and polyadenylation. Overall, RNA modifications dynamically reshape the structural landscape of RNA, enabling diverse functional interactions and responses, while advanced probing techniques like PARS and SHAPEseq continue to elucidate the global impact of these modifications.

Br{\"u}mmer~\textit{et~al.} found that A-to-I editing significantly enhances RNA secondary structures by reducing the accessibility of microRNA (miRNA) target sites~\cite{brummer2017structure}. This stabilization arises from the incorporation of inosine, which modifies the thermodynamic properties of RNA, leading to a more compact structure. Consequently, edited mRNAs exhibit reduced accessibility to Argonaute 2 (AGO2)-miRNAs, which typically bind to and destabilize unedited mRNAs. Importantly, A-to-I editing does not substantially alter the sequences of miRNA target sites; rather, it influences their accessibility through structural modifications. Experimental validation has shown that edited transcripts display higher expression levels than their unedited counterparts, underscoring the critical role of A-to-I editing in regulating mRNA stability and abundance through modulation of RNA secondary structures.

Furthermore, Boo and Kim recently emphasized the emerging role of RNA modifications in the regulation of mRNA stability, including m6A, N6,2’-O-dimethyladenosine (m6Am), 8-oxo-7,8-dihydroguanosine (8-oxoG), pseudouridine ($\Psi$), 5-methylcytidine (m5C), and N4-acetylcytidine (ac4C)~\cite{boo2020emerging}. Modifications such as m6A and its derivatives can either enhance or diminish mRNA stability, thereby affecting translation efficiency and degradation rates. These modifications can potentially alter the secondary and tertiary structures of mRNA, impacting the accessibility of RNA-binding proteins involved in mRNA surveillance and decay pathways. Consequently, specific RNA modifications can lead to either the stabilization or destabilization of mRNA, ultimately influencing protein synthesis levels and the overall expression of genes.

\subsection{RSS prediction with energy parameters for modified nucleotides}\label{sec:rss_modified_param}

As introduced earlier, the change in free energy associated with RSS folding is estimated using a set of empirical parameters from the nearest neighbor model, which is derived from optical melting experiments conducted on model systems in the wet lab.
Given the impact of RNA modifications on folding stability and the prevalence of m6A, Kierzek et al.~\cite{kierzek2022secondary} constructed a dedicated dataset that consists of a complete set of all nearest-neighbor parameters incorporating m6A in order to make modified RNA secondary structure prediction (Figure~\ref{MFEfig}\textbf{(B)} shows an example usage of this set). 
Similar to the Turner nearest neighbor parameters~\cite{mathews2004incorporating, mathews288expanded}, Kierzek et al. developed a full set of thermodynamic parameters for m6A as well as normal ACUG under the nearest neighbor free energy model.
Specifically, they estimated corresponding free energy changes at 37$^{\circ}$C with optical melting experiments of synthesized oligonucleotides.
After obtaining these new nearest neighbor parameters for m6A, they extended \textsc{RNAstructure}~\cite{reuter2010rnastructure} that we discussed earlier to incorporate m6A into the alphabet and use the new parameters to make RSS predictions.
This work showed promise in the accurate modeling of m6A-modified RNAs.
Notably, the authors also reported transcriptome-wide predictions with m6A, showing that methylation reduces the probability of adenosine being buried in helices (i.e., 21\% for A and 13\% for m6A), potentially driving widespread structural changes that influence RNA-protein interactions. 
Besides, the NNDB database~\cite{mittal2024nndb, turner2010nndb} (more details in the datasets section next) has also collected this set of nearest-neighbor parameters for RNAs with m6A modifications.

In a follow-up study, Szabat et al. conducted around 100 optical melting experiments in addition, focusing on m6A versus normal adenine to test and refine the corresponding free energy nearest neighbor parameters~\cite{szabat2022test}. 
Specifically, they utilized the RRACH motif, which is a known consensus motif of N$^6$-methylation in mammalian cells, with R representing a purine and H representing one of \{A, C, U\}. 
They experimented with the central site of RRACH, with or without methylation, in various secondary structure contexts, including helices, bulges, internal loops, dangling ends, and terminal mismatches.
With the m6A-expanded-nearest-neighbor parameters, the authors estimated the folding free energy changes, i.e., the folding stability, and compared them to the measured values from melting experiments.
As a result, the overall root mean squared deviation (RMSD) between experimental and predicted free energy changes across all experiments was 0.67 kcal/mol, indicating robust accuracy for the m6A-expanded parameters.
Moreover, the agreements between experimental and predicted folding free energy change with m6A were similar to those with normal A for most structural contexts.
The authors also revised the original parameters~\cite{kierzek2022secondary} under each structural context to further calibrate.
This work validates the potential RSS prediction capability for modified RNAs and provides a foundation to expand our understanding of m6A's roles in epitranscriptomic and gene regulations.
A later version update of RNAstructure also includes these revised parameters to its command line release to expand its functionality~\cite{Ali2023RNAstructure}.

Furthermore, as the most widely used RSS prediction tool, the ViennaRNA package also recently included support for modified RNA bases, starting from its version 2.6.0 update~\cite{varenyk2023modified}.
Instead of concentrating on one type of modification or experiments from one study, a comprehensive search was performed to identify available energy parameters from existing literature, which covers parameter sets like the m6A one discussed above~\cite{kierzek2022secondary, szabat2022test} and many more.
In total, the authors collected six different types of modifications from a number of experiments, including 7-deaza-adenonsine (7DA)~\cite{richardson2016nearest}, inosine~\cite{wright2007nearest, wright2018stability}, pseudouridine~\cite{hudson2013thermodynamic}, non-standard purine nucleotide nebularine~\cite{jolley2017loss}, dihydrouridine~\cite{dalluge1996conformational, chou2016blind}, as well as m6A~\cite{kierzek2022secondary}.
Then, the ViennaRNA package adopts a different strategy to accommodate for the effects of these modified bases by utilizing a hard- and soft-constraints framework~\cite{lorenz2016rna} to modify upon the RSS energy computations made with normal Turner's nearest neighbor parameters, only when there is a modified nucleotide and its corresponding parameters are available. 
This strategy waives the need for a huge multi-dimensional lookup table that is typically required to store a complete set of nearest neighbor parameters covering all possible cases, which is more memory efficient and better aligned with the sparse nature of the energy parameter data for modified bases.
Existing programs in the ViennaRNA package, like RNAfold~\cite{lorenz2011viennarna} etc., have been extended with this functionality.
At the time of writing, this work presents the largest set of RNA modification energy parameters for secondary structure prediction.

\section{Datasets}\label{sec: data}

\begin{ThreePartTable}
    \begin{TableNotes}\footnotesize
        \item \textbf{Note:} RNAStralign dataset is also commonly available in several follow-up studies with different versions (e.g., \href{https://zenodo.org/records/4430150}{\textsc{MXfold2} data}, \href{https://github.com/ml4bio/e2efold}{\textsc{E2Efold} data}, \href{https://huggingface.co/datasets/rouskinlab/RNAstralign}{HuggingFace}). 
    \end{TableNotes}
    
    \begin{small}    
    \begin{longtable}{lp{0.13\textwidth}p{0.5\textwidth}p{0.09\textwidth}}
    \caption{A summary of representative datasets discussed in this review.
    }
    \label{datasets}\\
    \hline
    \textbf{} & \textbf{Type} & \textbf{Data description} & \textbf{URLs} \\
    \hline
    \endfirsthead
    \multicolumn{4}{l}{Continuation of table}\\
    \hline
    \textbf{} & \textbf{Size} & \textbf{Data description} & \textbf{URLs} \\
    \hline
    \endhead
    \hline
    \multicolumn{4}{r}{More on next page}\\
    \endfoot
    \hline
    \insertTableNotes
    \endlastfoot
    \textbf{ArchiveII}~\cite{sloma2016archive} & RSS & Contains 3 975 seqeunces and structures ranging from 10 RNA families. & \href{https://rna.urmc.rochester.edu/publications.html}{Website} \\
    \textbf{bpRNA-1m}~\cite{danaee2018bprna} & RSS & Contains 102 318 sequences and structures of approximately 2500 RNA families, mainly collected from Rfam 12.2. Does not remove redundant sequences, but a processed subset, bpRNA-1m(90), does. & \href{https://bprna.cgrb.oregonstate.edu/index.html}{Website} \\
    \textbf{Rfam}~\cite{ontiveros2024rfam} & RSS & A comprehensive collection consisting of 90 190 RNA sequences of $>$4000 RNA families and all RNA types. Various sequence lengths from $<$20 short miRNAs to several thousands-nt-long lncRNAs. & \href{https://rfam.org/}{Website} \\
    \textbf{RNAStralign}~\cite{tan2017turbofold} & RSS & Contains 30 451 sequences from 8 RNA families. Sequence length ranges from 30 to 1800+. & \href{http://rna.urmc.rochester.edu/}{Website} \\
    \textbf{TORNADO data}~\cite{rivas2012tornado} &  & In total, contains 5 387 sequences. Contains two different collections A (sourced from literature, 11 families) and B (from Rfam, 22 families), each further splitting to TrainSet and TestSet. RNAs in A have longer lengths (10 to 700+) than in B (27 to 200+), $<$70\% sequence identity, and different structure distributions. & \href{https://rnajournal.cshlp.org/content/18/2/193/suppl/DC1}{Data link} \\
    \textbf{RNAcentral}~\cite{rnacentral2021rnacentral} & RSS & Contains comprehensive information about non-coding RNAs, including RNA sequences and structures, mRNA interactions, and RNA family classifications. & \href{https://rnacentral.org/}{Website} \\
    \textbf{NNDB}~\cite{mittal2024nndb, turner2010nndb} & Thermodynamic Parameters & Contains nearest neighbor parameters for normal and m6A modified base as well as DNA folding parameters, in the form of free energy changes, which is used for RSS predictions. & \href{https://rna.urmc.rochester.edu/NNDB/}{Data link} \\
    
    \textbf{PDB}~\cite{Berman2000} & 3D structure & Provides RNA 3D structures, 
    able to be used as a RNA folding benchmark, as was the case with Szikszai et al.~\cite{szikszai2024}.& \href{https://www.rcsb.org/}{Website} \\
    
    \textbf{MODOMICS}~\cite{cappannini2024modomics} & RNA Modification & Contains information about chemical structures, biological pathways, and sequences, enzymes of RNA modifications. Currently has more than 170 different RNA modifications, 429 different RNA modified residues (335 natural ones), and 1925 RNA sequences. & \href{https://genesilico.pl/modomics/}{Website} \\
    \textbf{ENCORI}~\cite{li2014starbase} & CLIP & Contains information about RNA-RNA and Protein-RNA interactions from CLIP-Seq data. Also has analysis of impact of interactions on gene expression in 32 cancer types. & \href{https://rnasysu.com/encori/}{Website} \\
    \textbf{eCLIP data}~\cite{van2016robust} & CLIP & Using the eCLIP protocol, currently has 119 RBPs and 102 RBPs interaction with RNA in the K562 and HepG2 cell lines respectively. & \href{https://www.encodeproject.org/}{Website} \\

    \hline
    \end{longtable}
    \end{small}
\end{ThreePartTable}

\subsection{RSS datasets}

\paragraph{ArchiveII}
ArchiveII~\cite{sloma2016archive} is an RNA secondary structure dataset compiled by Mathews Lab that is commonly used for the training and testing of RSS prediction methods containing a total of 3,975 RNA sequences and their structures. This dataset is an expansion of the previous RNA secondary structure dataset, also compiled by Mathews et al.~\cite{mathews1999expanded}, updating and including new structures. It contains the sequences and structures of 10 RNA families, such as small subunit ribosomal RNA, large subunit ribosomal RNA, 5S ribosomal RNA, Group I self-splicing introns, RNase P RNA, signal recognition particle RNA, tRNA, and tmRNA. The structural data was collected from databases such as RNA STRAND v2.0, 5S ribosomal RNA database, Rfam 9.1, and tmRDB. The dataset contains redundant sequences, or the same sequence of RNAs from different species, which have not been removed from the dataset. 

\paragraph{bpRNA-1m}
bpRNA-1m~\cite{danaee2018bprna} is another RNA secondary structure meta-database that has been compiled from multiple data sources, such as Comparative RNA Web (CRW)~\cite{cannone2002crw}, tmRNA database~\cite{andersen2006tmrdb}, tRNAdb~\cite{zwieb2003tmrdb}, Signal Recognition Particle database~\cite{rosenblad2003srpdb}, RNase P database~\cite{brown1998rpdb}, tRNAdb 2009 database~\cite{juhling2009trnadb}, and RCSB Protein Data Bank~\cite{rose2016rcsb}, and RFAM 12.2~\cite{nawrocki2015rfam}. bpRNA-1m is a comprehensive database that contains 102 318 RNA sequences and structures and approximately 2500 different RNA families. A less redundant version exists, bpRNA-1m(90), which removes sequences with greater than 90\% sequence similarity with at least 70\% alignment coverage. As a subset of bpRNA-1m, bpRNA-1m(90) contains fewer sequences or 28 370 sequences and structures. bpRNA-1m could also be used under an alias. MXFold2 uses this dataset and splits the dataset into three different subset datasets, TR0, VL0, and TS0, where they correspond to the training, validation, and testing datasets, respectively.

\paragraph{TORNADO dataset}
Rivas et al.~\cite{rivas2012tornado} curated a dataset in the development of their RSS prediction method, TORNADO. This dataset then serves as a benchmark widely used to measure the performance of a number of RSS prediction models later. The dataset consists of 4 different sets, TrainSetA, TestSetA, TrainSetB, TestSetB. TrainSetA and TestSetA were constructed by collecting sequences and structures from trusted literature. The TestSetA specifically ensures that there is no sequence redundancy by removing nearly identical sequences. This process results in a dataset with low sequence similarity, but as it is from 11 RNA families, the structural similarity is high. To combat this, TrainSetB and TestSetB were constructed by including 22 RNA families from the Rfam database. 

\paragraph{RNAStrAlign} RNAStrAlign~\cite{tan2017turbofold}, the most recent dataset by Mathews Lab, was constructed as a benchmark for the RSS folding algorithm, TurboFold II. The dataset consists of 8 different RNA families, 5S ribosomal RNA, Group I intron, tmRNA, tRNA, 16S ribosomal RNA, Signal Recognition Particle (SRP) RNA, RNase P RNA, and telomerase RNA, totaling 30 451 sequences, taken from disparate online databases. 

\paragraph{RNAcentral} 
While RNAcentral~\cite{rnacentral2021rnacentral} is a database that specifically contains data about non-coding RNA (ncRNA) sequences. The most recent version has started integrating and visualizing known RNA secondary structures of tRNA sequences, imported from GtRNAdb ~\cite{chan2016gtrnadb}. While the database does not provide a curated list of RNA secondary structures, it contains RNA secondary structure data of known ncRNA sequences, which can be used as a starting point for creating an RSS benchmark specialized in ncRNA folding.

\paragraph{Rfam} Rfam is a comprehensive RNA database that contains sequences and alignments from a wide variety of RNA families \cite{ontiveros2024rfam}. 
For each family, a seed multiple-sequence alignment is curated from a small set of representative sequences with a corresponding conserved secondary structure annotation. 
When such alignment and secondary structure annotation are not available from the literature, the Rfam team will generate them using programs like RNAalifold~\cite{bernhart2008rnaalifold} mentioned above, with manual adjustment.
The seed alignment is further used to build a covariance model, and subsequently, a full alignment with more sequences scored above a cutoff by the covariance model is added to the seed alignment. 
In general, Rfam is widely used in the field as the gold standard for training and assessing the accuracy of structure prediction programs.
At the time of writing, Rfam holds 4178 families across different species and for both coding and non-coding RNAs.

\paragraph{NNDB} Different from the other data reviewed above, the nearest neighbor database (NNDB)~\cite{turner2010nndb, mittal2024nndb}, as the name indicates, stores the nearest neighbor thermodynamic parameters for RNA and DNA from several experiments.
As the backbone of energy-based methods, these parameters are the key to RSS prediction and have been widely used, which is why we also include NNDB here.
Currently, the database contains both the 1999~\cite{mathews288expanded} and 2004~\cite{mathews2004incorporating} versions of the Turner's nearest neighbor parameters, the m6A modified parameters~\cite{kierzek2022secondary} as introduced earlier (section~\ref{sec:rss_modified_param}), and a set of DNA folding parameters.
For the three RNA parameter sets, free energy changes are stored as parameter lookup tables. 
The Turner 2004 version also has parameter tables for enthalpy changes.
In addition to providing the values of the parameters, NNDB also summarizes the rules and provides representative examples of how to use these parameters, which can be easily adapted to dynamic programming and other methods of development.

\paragraph{PDB data}
Although not secondary structure data, PDB-derived datasets may potentially provide indirect RSS information from the 3D structures. So, we still include them in this review for a brief discussion. Recent research demonstrates the usefulness of them in enhancing tertiary RNA structure prediction. RNA3DB, developed by Szikszai et al. \cite{szikszai2024}, utilizes PDB data to construct a comprehensive dataset optimized for training and evaluating deep learning models in RNA structure prediction. This dataset solves issues related to the restricted availability and diversity of experimentally determined RNA structures, providing a more solid basis for the development of computational tools. Additionally, the MARS and RNAcmap3 databases broaden the scope of RNA folding research by integrating RNA sequences from many sources to improve multiple sequence alignments, an essential process for precise secondary and tertiary structure predictions\cite{chen2024}. In total, these integrating PDB resources may provide potential source for addressing the data scarcity challenges associated with RNA secondary structure prediction.

\subsection{RSS related RNA modification  datasets}

\paragraph{MODOMICS} 
\href{https://genesilico.pl/modomics/}{\textsc{MODOMICS}}~\cite{cappannini2024modomics} is a comprehensive database that provides information about the chemical structure of RNA modifications, biological pathways, and RNA sequence location of modification, links to human diseases, and the participating RNA modification enzymes. This database collects information about more than 170 different types of RNA modifications that are currently still being discovered with the development of new high-throughput technologies. MODOMICs currently contains a total of 429 different RNA modified residues with 335 natural ones among them, from RNA types such as tRNA and small nucleolar RNA (snoRNA). It also hosts 1925 different RNA sequences from multiple RNA families such as tmRNA, tRNA, rRNA, small nuclear RNA (snRNA), snoRNA, and Piwi-interacting RNA (piRNA). There are many more data sources dedicated for RNA modifications. However, since this review is not for pure RNA modifications, we only include MODOMICS here as an representative of this class.

\paragraph{ENCORI} \href{https://rnasysu.com/encori/}{\textsc{ENCORI}}~\cite{li2014starbase}, also known as starBase v2.0, is a database that hosts detailed information about RNA-RNA and Protein-RNA interaction networks by analyzing crosslinking immunoprecipitation sequencing (CLIP-Seq) studies. 
Although not RNA modification nor RSS data directly, the data in ENCORI may potentially provide indirect information to link RNA modification and RSS folding, through relevant protein-RNA interactions for example. 
So, we put it here as an representative.
At the time of this writing, ENCORI has analyzed and hosted interactions from 2725 CLIP-Seq datasets, resulting in millions of miRNA, RBP, RNA interactions with ncRNA and mRNA. In addition to the study of miRNAs' impact on mRNAs through CLIP-Seq datasets, due to the advancements of Degradome sequencing, ENCORI also hosts analyses of miRNA-RNA interactions through the degradome-seq datasets. ENCORI also allows its users to study the effects of these RNA-RNA and protein-RNA interactions on gene expression in 32 different cancer types.  

\paragraph{eCLIP data in ENCODE}
eCLIP data ~\cite{van2016robust} utilizes a more efficient and robust method of CLIP-Seq, called enhanced CLIP (eCLIP). 
We include this dataset here for the same reason as for ENCORI.
With more efficient sample requirements and while retaining single-nucleotide resolution, eCLIP reduces the high failure rates of previous CLIP protocols. To find RBP-RNA interactions, eCLIP experiments were carried out on two human cell lines, K562 and HepG2. For the K562 cells, 119 RBPs were studied, while for the HepG2 cells, 102 RBPs were studied. The eCLIP dataset is publicly available on the \href{https://www.encodeproject.org/}{ENCODE} project website.

\section{Challenges and opportunities}
Predicting RNA secondary structures is challenging due to several key factors. 
First, there are significantly fewer known RNA structures compared to protein structures. 
This lack of data makes it hard to train machine learning models effectively, leading to biased results and lower prediction accuracy. 
Alternative rule-based dynamic programming algorithms are widely used but suffer from several issues, like the scalability for long sequences, etc.
Second, pseudoknots and noncanonical interactions also introduce further difficulties, as many traditional DP-based methods struggle to account for them. 
Third, limited thermodynamic parameters and challenges in sampling different RNA shapes add to the problem. 
Finally, RNA folding is a complex process. 
The secondary structures can significantly affect the tertiary structures, making prediction more complicated. 
These issues highlight the urgent need for better models and techniques in RNA secondary structure prediction.

Recent advancements in machine learning, especially those inspired by deep learning and natural language processing, greatly improve the prediction of RNA secondary structures. These techniques can use large RNA sequence datasets to boost prediction accuracy. Combining machine learning with traditional thermodynamic and physics-based models can help us better understand how RNA folds. The rise of RNA language models, such as RNAErnie~\cite{wang2024multi}, RNA-FM~\cite{chen2022interpretable}, and UNI-RNA~\cite{wang2023uni}, along with experimental data like chemical probing results~\cite{deigan2009accurate}, provides valuable resources for refining prediction models, particularly for longer RNA sequences. Collaboration between theoretical and experimental researchers can spark innovation. Moreover, considering environmental factors such as temperature, ligands, and ions can lead to predictions that are more relevant to biological contexts. These opportunities point to a promising future for RNA secondary structure prediction, with important implications for computational methods and biological insights.

Moreover, RNA secondary structure refers to the arrangement of base pairs formed through hydrogen bonding between nucleotides, which plays a crucial role in determining the tertiary structure and functionality of RNA molecules. In the context of mRNA vaccines, modifications such as converting uridine residues to N1-methylpseudouridine (m1$\Psi$) are strategically employed to enhance RNA stability and reduce immunogenicity, both of which are vital for vaccine efficacy~\cite{kariko2005suppression}. These modifications can significantly alter the free energy parameters and base-pairing interactions, thereby complicating the prediction of secondary structures. Therefore, developing RNA secondary structure prediction methods that accurately account for these modifications is essential for advancing RNA drug discovery and therapeutic applications~\cite{tanzer2019rna}. Accurate predictions enable the rational design of RNA molecules with specific properties, ultimately improving the effectiveness of RNA-based therapeutics. Consequently, integrating advanced prediction methodologies is critical for optimizing the therapeutic potential of RNA technologies.

RNA secondary structure prediction can be employed to understand the regulatory mechanisms of stress response and virulence in foodborne pathogens. For instance, RNA thermometers are non-coding RNA elements that regulate gene expression in response to temperature changes \cite{johansson2002rna}, playing a crucial role in the virulence of pathogens like \textit{Listeria monocytogenes} \cite{hanes2023protein}, \textit{Escherichia coli}\cite{zhang2022discovery}and \textit{Salmonella Typhimurium}\cite{liu2024comparative}. Additionally, small RNAs (sRNAs) have been shown to influence the expression of virulence factors in foodborne pathogens, with studies indicating their involvement in stress response and pathogenicity \cite{padalon2008small}. Furthermore, RNA secondary structure prediction is critical in developing gene therapy.  The RSS tools help identify structural elements influencing RNA stability, translation efficiency, and interactions with RBPs. For example, RBPs such as OAS proteins can inhibit gene expression by binding to specific secondary structures within therapeutic RNA molecules, thereby reducing their therapeutic efficacy \cite{zhai2023investigation}. By using RNA secondary structure prediction, researchers can modify RNA molecules to prevent such inhibitory interactions with RBP binding sites \cite{roundtree2017dynamic}. Additionally, RNA modifications, such as m6A, $\Psi$, or m5C, can improve the efficiency of gene therapy by enhancing RNA stability, reducing immune activation, and improving translational output \cite{zaccara2019reading}.

Lastly, in the context of RNA tertiary structure prediction, recent developments such as AlphaFold3 have garnered significant attention due to their ability to model heterogeneous macromolecular systems, including large RNA molecules. A recent study\cite{alphafold3RNA2024}, Structure Prediction of Large RNAs with AlphaFold3 Highlights its Capabilities and Limitations, provides a comprehensive assessment of AlphaFold3's performance on RNA structures of up to 5000 nucleotides. The study highlights both the potential and the limitations of this tool, particularly for predicting large RNA molecules whose experimental dimensions are known. While AlphaFold3 can generate plausible models, challenges persist, including severe steric clashes, occasional breaks in the phosphodiester backbone, and excessive sphericalization of structures, with this effect becoming more pronounced as RNA length increases. Notably, hydrodynamic radii calculated from AlphaFold3 models are substantially larger than experimental measurements under low-salt conditions but align more closely with experimental results in the presence of polyvalent cations. These findings suggest that while AlphaFold3 can be used for RNA structure prediction, especially for RNAs up to 2000 nucleotides, it may be required to identify geometrically accurate predictions free of structural artifacts. These limitations suggest that AlphaFold3 provides a useful starting point for RNA structure modeling; nonetheless, it requires further optimization and complementary approaches, such as experimental data integration or RNA-specific prediction tools, to achieve reliable RNA structure predictions. The static nature of PDB structure data, which captures only a single RNA conformation, also presents limitations to AlphaFold3.

\section{Conclusion}
\label{sec:conclusion}
In this review, we explored the advances in RNA secondary structure predictions, RNA modifications, and their interplays, highlighting the progression of the methodology and the connection between RNA structure and modification. 
RNA secondary structure prediction methods have evolved drastically over the past decades, moving from dynamic programming algorithms to sophisticated learning-based approaches that account for complex structure patterns such as pseudoknots and long-range interactions. 
Meanwhile, advances in RNA modification prediction tools have leveraged the progress in experimental data and various machine learning and deep learning paradigms to analyze the functional roles of many modification types.
The integration of RNA secondary structural information into modification prediction models, and vice versa, has expanded our understanding of the critical role of RNA in regulating gene expression, RNA-protein interactions, and other important biological processes.
Nevertheless, significant challenges remain in the field, such as the scarcity of data compared to the related field of protein structures. 
Addressing them will not only supply more accurate prediction methods but also pave the way for novel therapeutic applications and breakthroughs in quantitative biology.

\section*{Acknowledgments}
This work was supported in part by the the NIH grants P30 AG073105, U01 AG066833, U01 AG068057, U19 AG074879, and R01 AG071470.

\makeatletter
\renewcommand{\@biblabel}[1]{\hfill #1.}
\makeatother

\small
\bibliographystyle{vancouver}
\bibliography{main}  

\end{document}